\documentclass[12pt]{article}         
\usepackage{graphicx}
\usepackage{mathptmx}     
\usepackage[round]{natbib}
\usepackage[T1]{fontenc}
\usepackage[utf8]{inputenc}
\usepackage{amsmath}
\usepackage{amssymb}
\usepackage{graphicx}

\usepackage[german,english]{babel}
\usepackage[noabbrev]{cleveref}
\usepackage{xfrac}

\begin{document}

\title{Einstein on Involutions in  Projective Geometry}

\author{Tilman Sauer \; and \; Tobias Schütz\\[1cm] .
{\small   Institute of Mathematics} \\[-0.1cm]
{\small   Johannes Gutenberg University Mainz} \\[-0.1cm]
{\small   D-55099 Mainz, Germany}\\[-0.1cm]
{\small   Email: tsauer@uni-mainz.de, tschuetz@uni-mainz.de} }          

\bigskip

\date{Version of \today}

\maketitle

\begin{abstract}
We discuss Einstein's knowledge of projective geometry. We show that two pages of Einstein's Scratch Notebook from around 1912 with geometrical sketches can directly be associated with similar sketches in manuscript pages dating from his Princeton years. By this correspondence, we show that the sketches are all related to a common theme, the discussion of involution in a projective geometry setting with particular emphasis on the infinite point. We offer a conjecture as to the probable purpose of these geometric considerations.
\end{abstract}

\section{Introduction}

Assessment of Einstein's mathematical knowledge and proficiency is crucial for a historical understanding of his creativity as a physicist. The question of Einstein's knowledge of mathematics has therefore been discussed frequently in the historical literature.
\citet{PyensonL1980Education} already looked closer at Einstein's mathematics education, concluding that he had an ``excellent preparation for his future career'' (p.~399). The publication of the first volume of Einstein's \emph{Collected Papers} in 1987 then provided the most important primary documents \citep{cpae1}. 
Publication of the early correspondence between Einstein and Mileva Mari\'c not only provided details about Einstein's undergraduate studies in Zurich but also revealed the close communication between Einstein and his fellow student and later wife. The collaboration between the two gave rise to a much-debated issue of Mileva Mari\'c's role in their collaboration based on conjectures that she had provided the mathematics for Einstein's discovery of special relativity  \citep{TroemelPloetzS1990Mileva,StachelJ1996Einstein}.
\citet{CahanD2000Education} and more recently \citet{BraccoC2017Quand} also look at Einstein's early education but focus on his physics training. The question plays a role for an assessment of Einstein's heuristics as well as the role of mathematics in Einstein's physical theorizing \citep{ZaharE1980Einstein}, \citep{NortonJ2000Nature}. 
Studies of Einstein's own calculations in the so-called Zurich Notebook from 1912 provided insights into his own learning process with respect to tensor calculus as well as into his way of putting mathematical concepts to use in his attempts to find a relativistic field theory of gravitation \citep{NortonJ1984Einstein,RennJEtAl1999Heuristics,janssen_norton_renn_sauer_stachel_2007}.

Einstein's own published writings do discuss mathematical topics although rarely ever with explicit references to specialized literature. In his collaboration, he often relied on mathematicians to assist him in his physical theorizing. A well-known example is the 1913 \emph{Entwurf} paper, presenting a precursor version of general relativity, which was co-authored with his mathematician friend Marcel Grossmann  \citep{EinsteinAEtal1913EntwurfZ}. The paper clearly declares their respective division of labor by the fact that Einstein authored its physical part and Grossmann its mathematical part. The final breakthrough to the fully covariant general theory, however, Einstein made on his own in late 1915 in competition with the mathematician David Hilbert.

In this paper, we intend to shed more light on Einstein's mathematical knowledge, understanding and skills by looking at Einstein's own research notes. We will discuss some specific arguments in projective geometry that Einstein is documented to have pondered in his deliberations.

In this project, we rely on the analysis of unpublished research notes and calculations, in particular his unpublished Princeton working sheets \citep{sauer_2019}. It turns out that Einstein very rarely used graphical sketches or geometric constructions in his notes. This in itself is a noteworthy fact in light of his own statements about the non-verbal character of his thinking \citep[Appendix II]{HadamardJ1945Psychology}.
There is the occasional sketch of a coordinate system or reference frame. But in contrast to, for example, notes by his friend Grossmann, he rarely prepared any more elaborate geometrical constructions or the like. Instead, the vast majority of his private notes and calculations are algebraic in nature. We did identify, however, some pages in the Einstein Archives which document some rather specific arguments in the context of projective geometry. We will discuss these pages in some detail and claim that the very same argument was entertained by Einstein both in a research notebook dated to the years 1912--1915, and in research notes which we have dated to the summer of 1938.

The outline of our paper is as follows.
In section \ref{sec:studies}, we recapitulate the basic facts about Einstein's studies and his academic training in mathematics, establishing that he was probably well acquainted with basic concepts of projective geometry of the plane. In section \ref{sec:evidence}, we then introduce our primary evidence, a double page in the Scratch Notebook of his Prague and Zurich years as well as a few manuscript pages from his Princeton working sheets. We will argue that these pages contain geometric constructions as well as analytical calculations that all pertain to the notion of projective involutions. A brief summary of our reconstruction of the double page of the notebook is given in section \ref{sec:summary}.
In section \ref{sec:literature}, we discuss some relevant literature on projective geometry, both to the extent that Einstein may have been familiar with it and with respect to discussions of specific aspects of the reconstruction.
In the following sections \ref{sec:inv}, \ref{sec:pas}, and \ref{sec:cal}, we will show in detail how very similar considerations are underlying a number of sketches found in the Princeton manuscripts.
In section \ref{sec:inv}, we discuss involutions on a line, section \ref{sec:pas} discusses involutions on a conic. We show how Einstein constructs geometric representations of involutions on a conic by starting from Pascal's theorem and reducing the hexagon to a quadrangle.
In section \ref{sec:cal}, we will give a reconstruction of a sketch and calculation in which Einstein gives both an analytical treatment and a graphical representation in terms of radical axis and involution by means of intersecting circles. We show that the corresponding consideration of the Scratch Notebook is a special case of the more general argument on the later manuscript page.
For the sake of completeness, we mention and briefly discuss some further pertinent sketches in section \ref{sec:further}, and
section \ref{sec:alternatives} serves as a caveat by pointing out some alternative scenarios of interpretation.
In section \ref{sec:conjecture}, we finally address the question of the overall context of Einstein's considerations of involution. We formulate a conjecture as to the probable purpose of Einstein's considerations,
We end in section \ref{sec:conclusion} with some concluding remarks.

\section{Einstein's Studies in Projective Geometry}
\label{sec:studies}

\subsection{Early Studies in Munich, Milano, and Aarau}

Einstein attended primary school and spent the first years of secondary school in Munich, first at a local \emph{Volksschule}, then at the renowned \emph{Luitpold-Gymnasium}.\footnote{For the curricula of Munich Volksschule and of Luitpold-Gymnasium, see \citep[Appendixes A and B]{cpae1}.} During those years, he also benefitted from private tutoring by the medical student Max Talmey who exposed him to a variety of advanced texts in all fields. In his autobiographical recollections, he recalled being fascinated by a magnetic compass and then continued:
\begin{quote}
At the age of 12 I experienced a second wonder [...]: in a little book dealing with Euclidean plane geometry. 
\citep[p.~9]{EinsteinA1949Autobiographisches}
\end{quote}
The book which he also called the ``holy geometry booklet'' (ibid., p.~11) most likely was the second part of Sickenberger's (\citeyear{SickenbergerA1888Leitfaden}) \emph{Leitfaden der elementaren Mathematik}, dealing with \emph{Planimetrie}. On Talmey's advice, he probably also studied Spieker's (\citeyear{SpiekerT1890Lehrbuch}) \emph{Lehrbuch der ebenen Geometrie}, see \citep[p.~lxi]{cpae1}. One of his earliest known writings is a comment in the margin of \citet{HeisEschweiler1881Lehrbuch}, criticizing a proof that the cylinder can be developped into a plane \citep[p.~3]{cpae1}.
He also remembered that he was studying mathematics eagerly as a teenager:
\begin{quote}
At the age of 12--16 I familiarized myself with the elements of mathematics together with the principles  of differential and integral calculus.
\citep[p.~15]{EinsteinA1949Autobiographisches}
\end{quote}
After Einstein had left Munich, he spent a year with his family in Milano engaging in autodidactic studies in preparation for an entrance examination to be admitted as a student at the Zurich Polytechnic. Failing this  examination, he was nonetheless encouraged and advised to finish his secondary schooling. Accordingly, he attended then the final class at Aargau cantonal school from October 1895 to September 1896. One of his teachers there was Heinrich Ganter who taught geometry following his own textbook. It was co-authored with Ferdinand Rudio, professor at the Polytechnic \citep{GanterRudio1888Elemente}, and treated plane geometry with an extensive discussion of conic sections.\footnote{See \citep[p.359]{cpae1}, we thank Klaus Volkert for pointing out the reference to \citep{GanterRudio1888Elemente}.} Einstein's school record from the Aargau cantonal school indicates that he received the highest grades in algebra and geometry.\footnote{See the Aargau cantonal school record and the final grades in \citep[Docs.~10, 19]{cpae1}. Incidentally, the grading system at the school changed for the school year from the German system (1 is highest, 6 is lowest) to the Swiss system (6 is highest, 1 is lowest), resulting in a change of his grades from 1 to 6 in algebra and geometry \cite[p.16]{cpae1}, a feature that occasionally causes confusion about Einstein's mathematical skills as a student.}

One of the questions for his \emph{Matura} examinations in Aarau concerned the following problem of conic sections in plane Euclidean geometry:
\begin{quote}
Consider a circle of radius $r$, centered on the origin of a rectangular coordinate system. At each point along the $x$-axis, another circle is constructed, with center at this point and diameter determined by the intersections of the perpendicular to the $x$-axis with the original circle. The circles so constructed are enveloped by an ellipse of semiaxes $r$ and $r\sqrt{2}$. When the distance of the centers of the circles from the origin exceeds a certain maximum value, the circles cease to touch the envelope. Prove the last two statements and determine this maximum value. \cite[Doc. 23, n. 5]{cpae1}
\end{quote}
\begin{figure}
\centering
\includegraphics[width=0.68\textwidth]{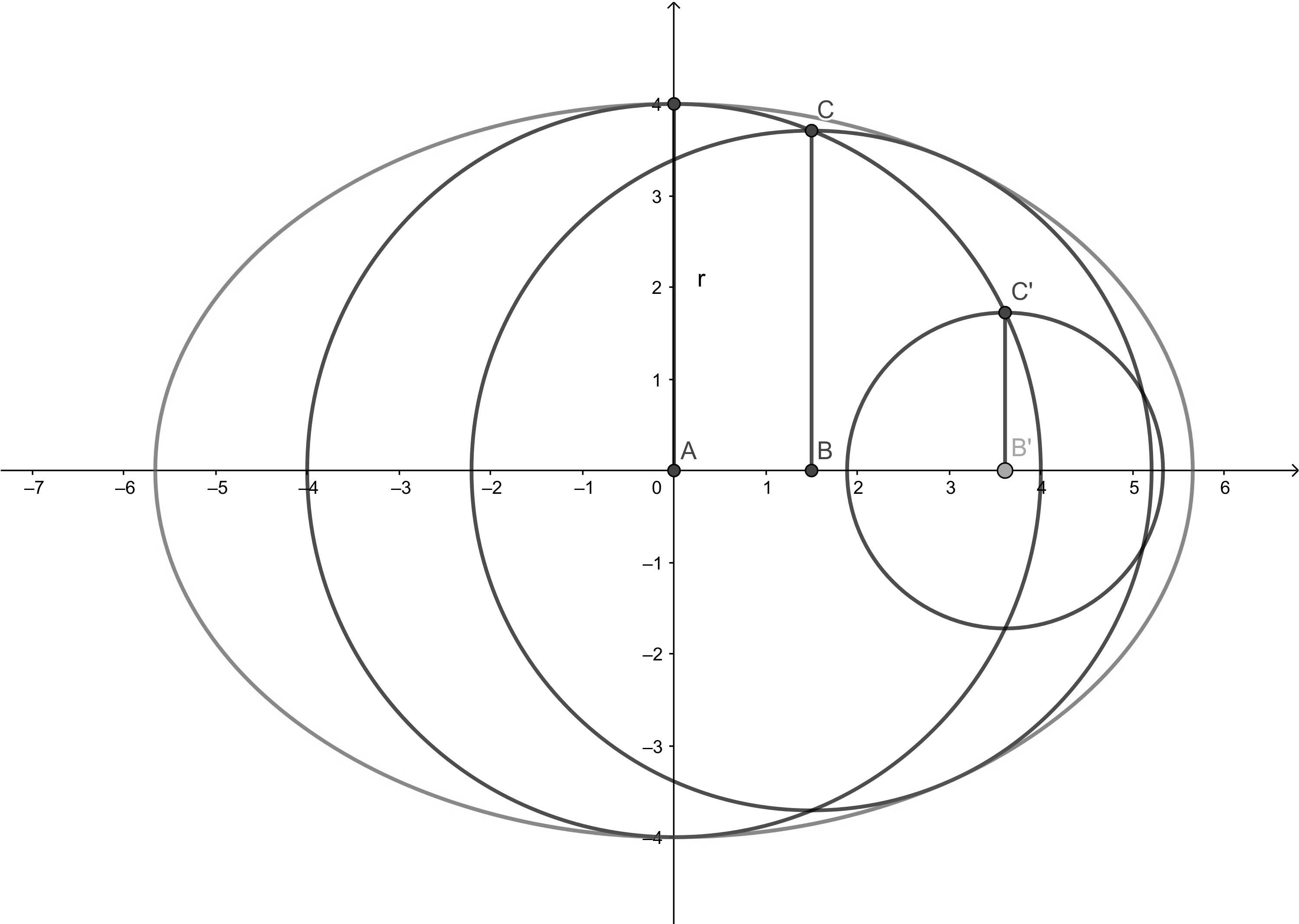}
\caption{Illustration of a geometry problem presented to Einstein for his \emph{Matura} examination.}	
\label{fig:ae-matura}
\end{figure}
For an illustration of this problem, see Fig.~\ref{fig:ae-matura}. In his solution, which did not involve any drawings, Einstein correctly derived the analytic expression for the enveloping ellipse and the distance $r/\sqrt{2}$ of the center from the center of the ellipse of the smallest osculating circle.\footnote{For more detailed discussion of Einstein's Aarau \emph{Matura}, see \citet{HunzikerH2005Einstein}.}

\subsection{ETH 1896 and 1897}
\label{subsec:eth}

After graduating from Aargau cantonal school, Einstein took his undergraduate studies in Switzerland at Zurich's \textit{Eidgenössische Polytechnische Schule} (Federal Polytechnic School, in 1911 renamed to \textit{Eidgenössische Technische Hochschule} (Swiss Federal Institute of Technology, ETH) from 1896 to 1900. He was one of eleven students initially enrolled in department VI A: the School for Mathematics and Science Teachers, section of mathematics, physics, and astronomy.\footnote{`Schule für Fachlehrer in mathematischer und naturwissenschaftlicher Richtung. Mathematisch-physikalische Sektion''  \cite[43]{cpae1}, see also \textit{ETH Reglement (1899)}, Albert Einstein Archives, Archival Number 74-593.} 

In his first year at the Polytechnic, Einstein took a class by Otto Wilhelm Fiedler on descriptive geometry, ``Darstellende Geometrie'', and in the second year, in summer 1897 and winter 1897/98, he attended lectures on projective geometry, ``Projektivische Geometrie~I'' and ``Projektivische Geometrie~II,'' respectively, also held by Fiedler. In a letter to Mileva Mari\'c, he wrote in 1898: ``Fiedler lectures on projective geometry, he is the same indelicate, rude man as before \& in addition sometimes opaque, but always witty \& profound -- in brief, a master but, unfortunately, also a terrible schoolmaster.''\footnote{``Fiedler liest projektivische Geometrie, derselbe undelikate, rohe Mensch wie früher \& dabei manchmal undurchsichtig, doch immer geistvoll \& tief -- kurz ein Meister aber leider auch ein arger Schulmeister.'' \cite[Doc. 39]{cpae1}} Fiedler's (\citeyear{FiedlerW1871Geometrie}) textbook on descriptive geometry already included many elements of what later became projective elements, and its third edition \citep{FiedlerW1883Geometrie,FiedlerW1885Geometrie,FiedlerW1888Geometrie} had not only expanded to three volumes but its title programmatically expressed an ``organic connection'' between descriptive geometry (``Darstellende Geometrie'') and projective geometry (``Geometrie der Lage'').

As to the actual content of Fiedler's lecture at the ETH in 1897,
the first of these lecture courses is documented by a transcript, written by Marcel \citet{grossmann_1897}. Grossmann was a good friend and classmate of Einstein and obtained his diploma together with Einstein in 1900. After this, he became assistant to Fiedler, obtained his Ph.D. in 1902 under Fiedler's supervision and was from 1907 on his successor for ``Descriptive Geometry'' (``Darstellende Geometrie'') und ``Projective Geometry'' (``Geometrie der Lage'') \cite[100--119]{Graf-Grossmann_2015}. 
We know from Einstein's recollections \cite[147]{Einstein_1955} and a letter to Grossmann that he frequently borrowed Grossmann's notes for test preparation, calling himself a ``slovenly guy who wouldn't even have been able to pass his examinations without the help of Grossmann's notebooks.''\footnote{``ein Schlamper [...], der ohne Hilfe von Grossmanns Heften nicht einmal seine Examina hätte machen können'' \cite[Doc. 226]{cpae14}.} In fact, in one of Grossmann's lecture transcripts of infinitesimal geometry \cite[105]{grossmann_1898}, there is a note not made by Grossmann that can most probably be traced back to Einstein \citep{sauer_2015}. It is safe to assume that Einstein carefully studied Grossmann's lecture notes on Fiedler's projective geometry, too.

Topics of the first lecture course in projective geometry of summer 1897 \citep{grossmann_1897} included, among others, the cross ratio of four distinct points, homogeneous coordinates, duality, conics, collineations, harmonic relations, and imaginary elements. Besides, Fiedler discussed involutions and Pascal's theorem. 
As listed in his Record and Grade Transcript, Einstein received grades $4.5$ and $4$ (of $6$) in the two courses of projective geometry, respectively.\footnote{See \textit{Matrikel. Einstein, Albert (1900)}, Albert Einstein Archives, Archival Number 71-539, \citep[Doc.~28]{cpae1}} 
Einstein's grades result in a final grade of $4.25$ of his final transcript (``Abgangszeugnis'') which was his worst one.\footnote{See \textit{Abgangszeugnis. Einstein, Albert (1900)}, Albert Einstein Archives, Archival Number 29-234.} Marcel Grossmann received in the two courses of projective geometry grades of $5$ and $5.5$.\footnote{See \textit{Abgangszeugnis. Grossmann, Marcel (1900)}, Albert Einstein Archives, Archival Number 70-755.}

It should be noted that prior to axiomatic foundations of the field which, to be sure, were fully worked out at the latest with  
\citet{VeblenYoung1910Geometry1,VeblenYoung1910Geometry2},
projective geometry was not a clearly defined field, and many topics that today are seen as characteristically projective  were discussed in the nineteenth and early twentieth century in other contexts as well, say, descriptive geometry. This is the case, in particular, with Fiedler's understanding of geometry \citep{VolkertK2021Fiedler}. %

Thus, as a student of the Aargau cantonal school and the Zurich Polytechnic, Einstein was exposed to an education in mathematics and physics of high standards, including training in geometry in its various subdisciplines. He was a good student with excellent grades in mathematics and physics. His academic teacher Fiedler was instrumental in creating projective geometry as a field of academic study \citep{KitzS2015Geometrie}, and we know that Einstein studied Fiedler's and other lectures, at least, through the carefully worked out notes prepared by his fellow student and friend Marcel Grossmann, who incidentally became Fielder's successor at the ETH in 1907. 

In order to assess to what extent Einstein retained an active command of the more sophisticated mathematical tools and concepts that he was exposed to during his studies, we need to turn to his later writings. By its very nature, a particularly revealing source are private research notes, in which he explored ideas for himself. Regarding projective geometry, we will here look at some of those manuscript pages.

\section{Evidence: the Scratch Notebook and the Princeton Manuscripts}
\label{sec:evidence}

\begin{figure}[htbp]
\centering
\includegraphics[width=.95\textwidth]{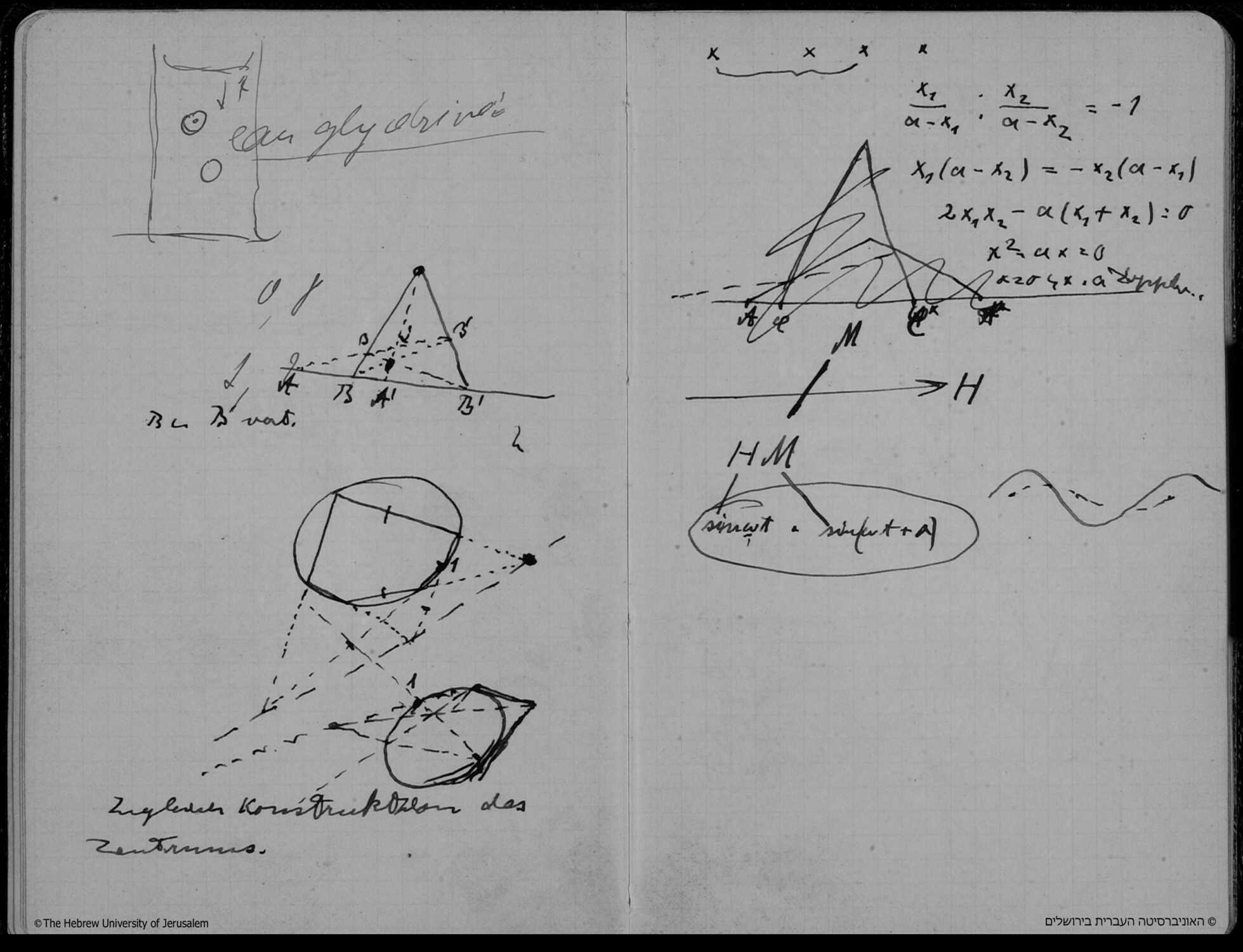}
\caption{Double page in Einstein's Scratch Notebook with sketches on projective geometry. AEA~3-013 [pp.~49--50], \citep[p.~588]{cpae3}. Reproduced with permission from the Albert Einstein Archives. \copyright The Hebrew University of Jerusalem, Israel. Digital image photographed by Ardon Bar Hama.}
\label{fig:notebook_total}
\end{figure}

Among Einstein's early research notes which are extant there is a Scratch Notebook with entries dated to the years 1910--1914.\footnote{AEA~3-013, \citep[Appendix A]{cpae3}. For more detailed discussion of various other entries in this notebook, see \citep{renn_sauer_stachel_1997}, \citep{janssenm1999rotation}, \citep{sauer_2008}, \citep{rowe_2011}, \citep{BuchwaldDEtal2013Note}. }
The double page of the Scratch Notebook that we will reconstruct in this article, is displayed in Fig.~\ref{fig:notebook_total}.
At the top of the left hand page, it displays a sketch drawn with pencil and the words ``eau glycerin\'e'' written in an unknown hand. In the middle of the right hand page, it shows sketches that carry letters $H$ and $M$ as well as the expression $\sin(\omega t) \cdot \sin(\omega t + \alpha)$.\footnote{The multiplication dot could also be a plus sign or an equal sign. For another possible context of this part of the page, see \citep[p.~359]{cpae3}.} We conjecture that these entries are related to the discussion of problems related to Brownian motion and to magnetism, respectively. They may reflect conversations Einstein had with Jean Perrin and Paul Langevin, both of whom he met on a visit to Paris to give a lecture on the law of photochemical equivalence to the Societ\'e de physique on 27 March 1913, see \citep[Doc.~437]{cpae5}. We will not be concerned with these entries on the double page but focus on the remaining entries. 

The double page is preceded in the notebook by a sequence of pages that contain calculations on gravitational lensing, dated to spring 1912. It is immediately followed by a double page that also contains a calculation on gravitational lensing, which, however, was written at a later date, probably in late 1915. For a detailed discussion of the surrounding pages and calculations, see \citet{sauer_2008,sauer_schuetz_2019} and references therein.

Interpreting the four $x$'s with an underbrace at the top of the right hand page as an independent sketch, we therefore find five sketches on this double page---three on the left hand side and two on the right hand side---as well as five lines of calculations that are related to projective geometry. It is our aim to account for these entries.

\begin{figure}[htbp]
\centering
\includegraphics[width=.49\textwidth]{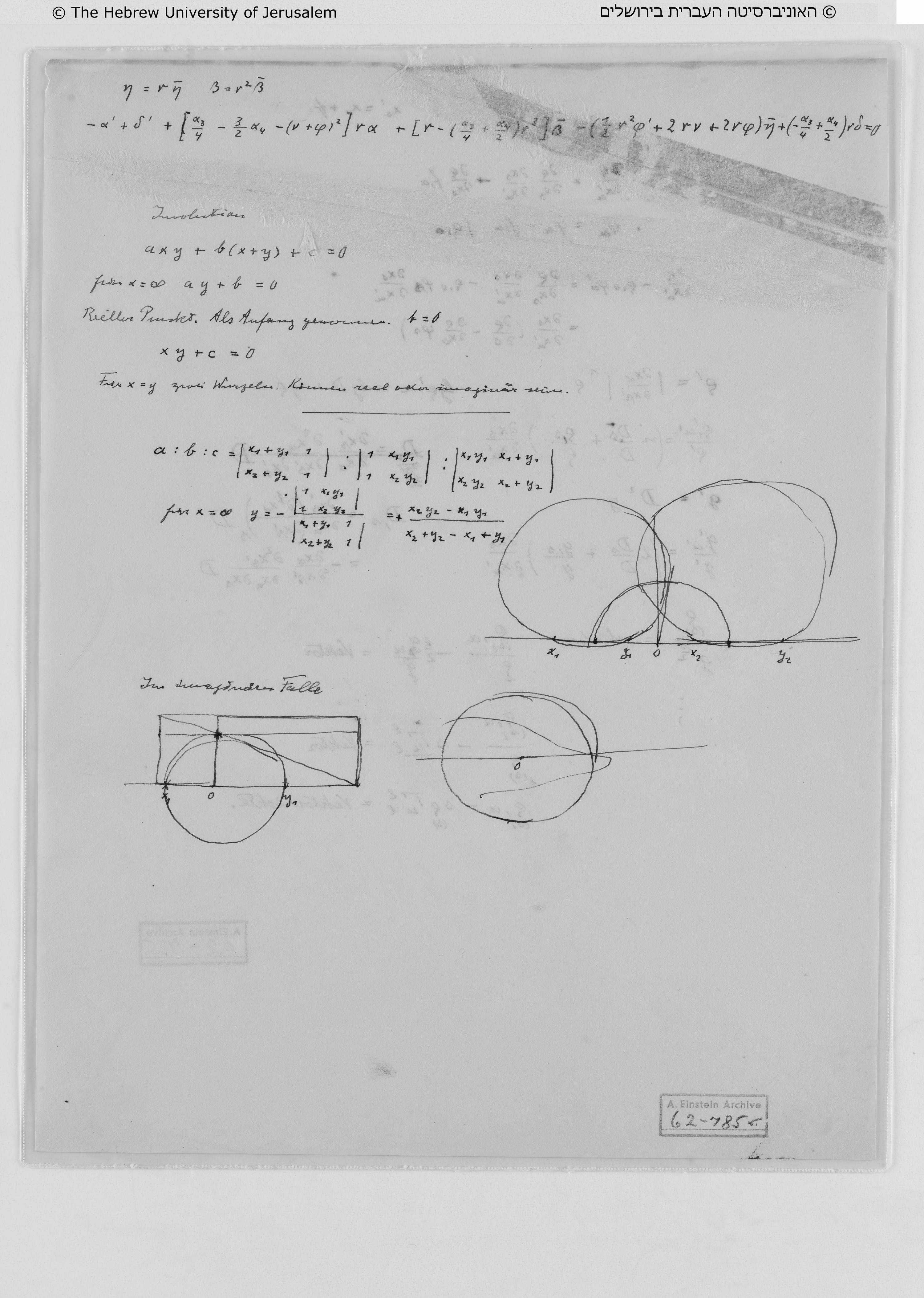}
\includegraphics[width=.49\textwidth]{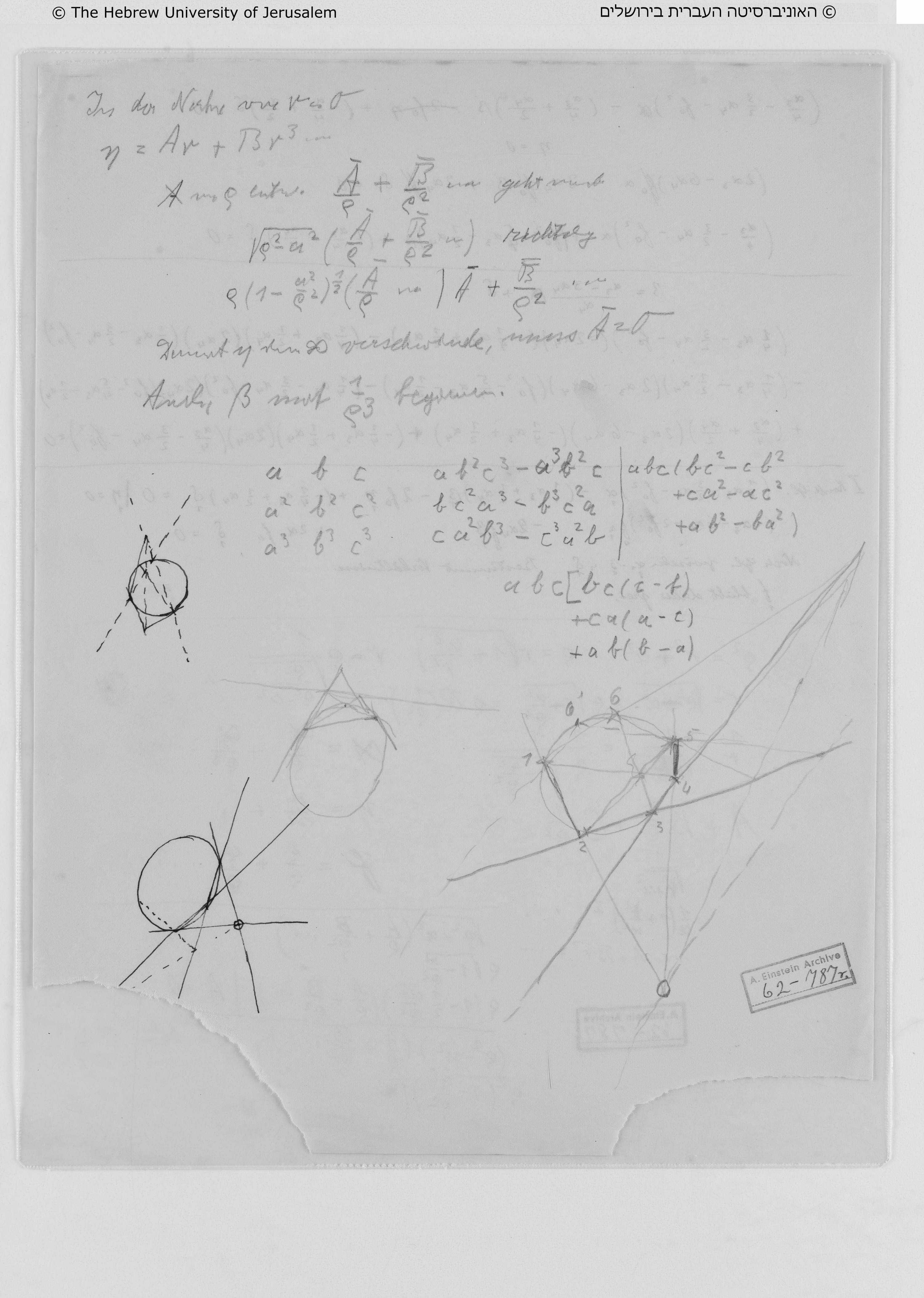}
\includegraphics[width=.49\textwidth]{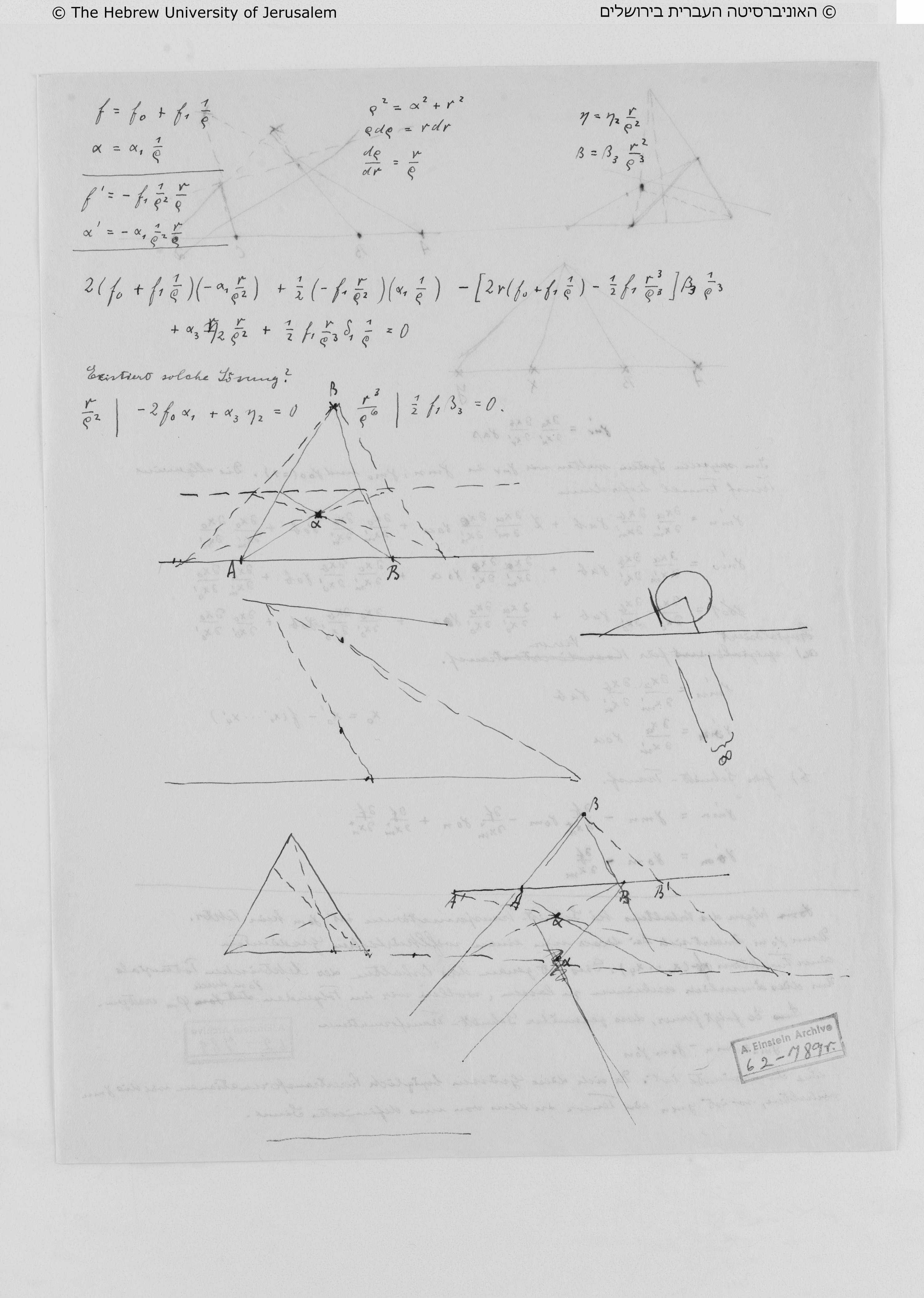}
\includegraphics[width=.49\textwidth]{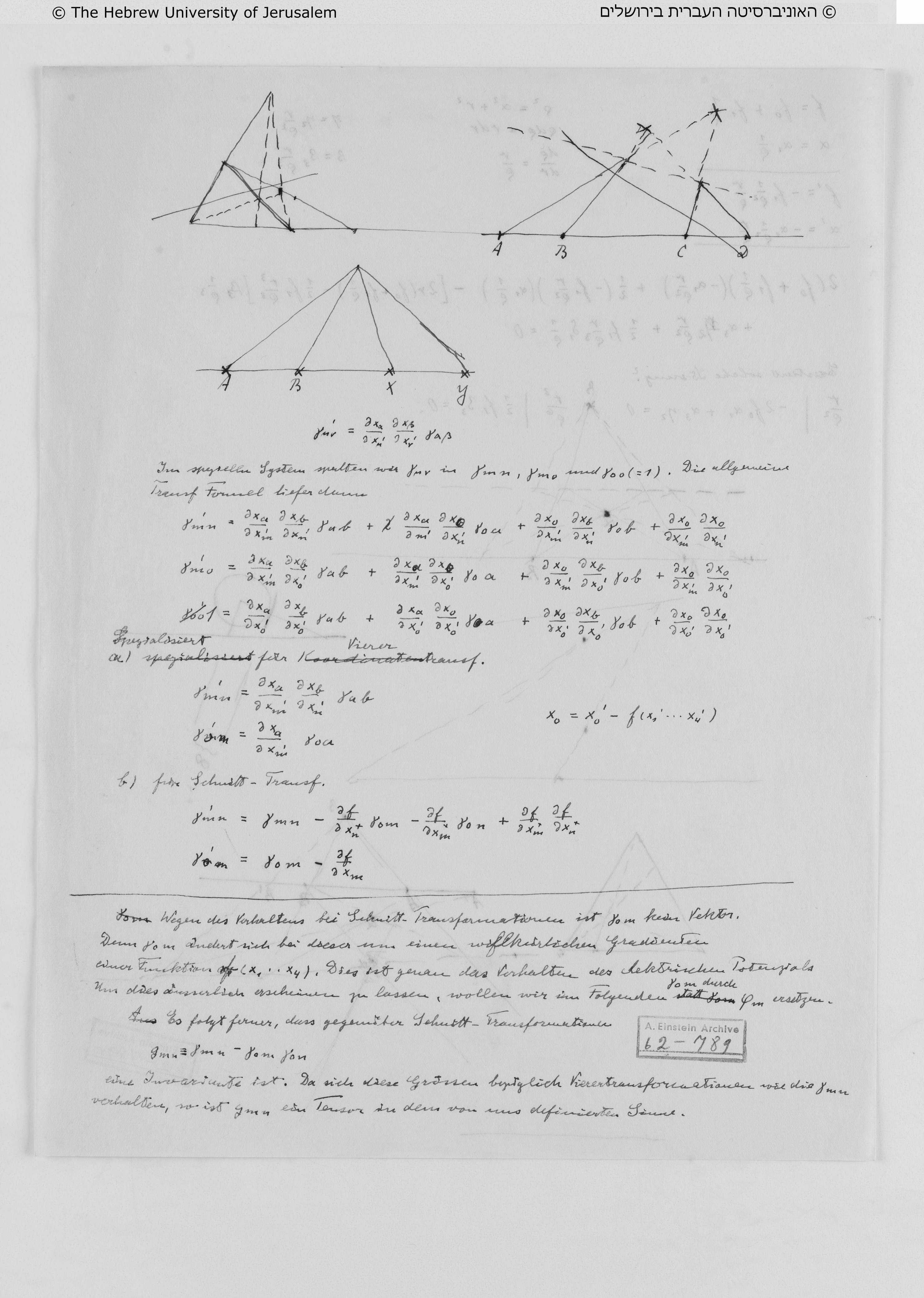}
\caption{Manuscript pages AEA~62-785r, 62-787r, 62-789, 62-789r. Reproduced with permission from the Albert Einstein Archives. \copyright The Hebrew University of Jerusalem, Israel. Digital image photographed by Ardon Bar Hama.}
\label{fig:62-785r-789r}
\end{figure}

Our reconstruction of the double page in the Scratch Notebook is based on a striking similarity that the sketches displayed there show with sketches that are found on three of Einstein's Working Sheets from his Princeton years.\footnote{For a discussion of these working sheets, see \citet{sauer_2019}.} The pages in question carry archival numbers AEA~62-785r, 62-787r, and 62-789 as well as 62-789r, see figure \ref{fig:62-785r-789r}. All four manuscript pages can unambiguously be dated to the period between June and August 1938, since surrounding calculations and phrases directly link these pages to correspondence between Einstein and Peter Bergmann from that time about their paper on a \emph{Generalization of Kaluza's Theory of Electricity}  \citep{einstein_bergmann_1938} as well as a related Einstein manuscript.\footnote{See \citet{SauerSchuetz2020Manuscript}. A detailed discussion will be given elsewhere.} While the relative sequence of the three sheets is unclear, it is very likely that entries on the recto 62-789r precede entries on the verso 62-789.

The close connection between these sketches is in itself an intriguing fact. If the similarity were to result from Einstein's entertaining the very same idea, it would be another instance of a long period of time between two characteristic trains of thought. Similarly, the surrounding lensing calculations of the Scratch Notebook from 1912--1915 were rediscovered, as it were, by Einstein in 1936 \citep{renn_sauer_stachel_1997}.

In addition to these pages, we will consider two more archival items that carry related sketches. One is an undated manuscript page AEA~124-446, likely from the early 1930's.\footnote{The Einstein Archives date the letter by provenance to the time period in which Einstein stayed in his summer house in Caputh, i.e. 1929--1933.} The other is AEA~6-250, a page from the Einstein-Bergmann correspondence, dated to the end of June 1938.\footnote{Again, Einstein’s letter AEA~6-250 is undated, but can be dated to between 21 and 30 June 1938 by surrounding correspondence.
%but refers to Bergmann’s previous letter AEA~6-247 written on 20 June. Since Einstein wrote another letter AEA~6-249 on June 21, referring to another letter AEA~6-246 Bergmann’s from 18 June, but not mentioning the letter AEA~6-247 from 20 June, we argue that AEA~6-250 was written after 21 June. Bergmann’s reply AEA~6-251 was written on 30 June, such that Einstein must have written AEA~6-250 before 30 June.
}

\section{Summary of our reconstruction}
\label{sec:summary}

Despite the apparently disparate character of the entries on the double page, we will show that all five sketches as well as the five lines of calculations are related to each other as a cluster of considerations revolving around the notion of involution in a projective geometry setting. In summary, our reconstruction of the page is the following. 

In the first sketch, Einstein constructs an example of a hyperbolic involution on a line by means of two consecutive perspectivities. The involution interchanges points $B$ and $B'$ while points $A$ and $A'$ are the two uniquely determined fixed points. In similar sketches from the Princeton manuscripts, he also investigated the case where one of the points of the involution moves to infinity.

The next sketch shows a construction of Pascal's theorem for a hexagon inscribed in a conic. The purpose of this construction appears to be to provide a basis for the geometric construction of an involution in a conic by means of an inscribed quadrangle. He obtains the third sketch on the page by letting two pairs of adjacent points of the hexagon approach each other, so that the hexagon becomes a quadrangle. The respective secants of the conic are marked and become tangents after this transformation. In order to arrive at a hyperbolic involution, he also moves the point marked 1 along the conic across the tangent point. The procedure amounts to the geometric construction of a hyperbolic involution in the conic. The very same procedure of starting from an inscribed hexagon to an inscribed quadrangle is found with more explicit notation on manuscript page 62-787r. In addition to the case of a hyperbolic involution, Einstein there also draws two additional situations which represent the cases of elliptic and parabolic involution. On that page, he also considers the special case where two of the opposite sides of the inscribed hexagon for Pascal's theorem become parallel. The Pascal line then becomes parallel to the sides and one of the three points on the Pascal line moves to infinity.

At the top of the next page, Einstein again considers involution on a line. He draws four collinear points and marks two of them which we interpret as indicating the two invariant points. Underneath we find a sketch that he deleted which we, again, interpret as a construction of hyperbolic involution on a line, this time, however, by means of a complete quadrangle.

The remaining entries to be accounted for are five lines of calculation. We interpret these as a special case of the general equation of involution for coordinates $x_1$ and $x_2$, which are then postulated to be invariant points of the involution, yielding two solutions $x=0$ and $x=a$. A corresponding and much more explicit calculation along these lines is found on manuscript page 62-785r. That calculation was explicitly headed by the term ``Involution.'' It starts from the general equation of involution and specifies to the case of one infinite point. In addition, the calculation on the manuscript page is accompanied by several sketches that give geometric constructions of the corresponding involution by means of the concept of a radical axis. Here again all three cases of hyperbolic, elliptic and parabolic involution are being considered.

Our reconstruction heavily depends on the similarity of the corresponding entries in the Princeton manuscript pages and the translatability between the notebook entries and the Princeton manuscript pages. In sections \ref{sec:inv}, \ref{sec:pas}, and \ref{sec:cal}, we will provide a detailed reconstruction of these connections. They are also visualized by some animations which illustrate the processes of construction and transformation that are involved in the reconstruction. These animations are provided as supplementary material to this article.\footnote{The animations will be made available as supplementary material to the published version of this paper.}

\section{Projective Geometry}
\label{sec:literature}

Before entering into our detailed reconstruction, we briefly mention some of the available textbook literature on projective geometry, both by texts that were available to Einstein as well as more modern treatises. We have not been able to identify any single work which might have triggered, or would have been an obvious source for, Einstein's sketches. None of the geometric theorems and properties that underlie Einstein's sketches are particularly specific to any single author, let alone original to Einstein. As a consequence similar sketches can be found in many places in the literature. 

Beginning with sources that would have been available to Einstein prior to 1938, we mention again Fiedler's textbook on Descriptive Geometry \citep{FiedlerW1871Geometrie} or \citep{FiedlerW1883Geometrie,FiedlerW1885Geometrie, FiedlerW1888Geometrie} and his Lectures, as well as Grossmann's Lectures. As noted before, from an axiomatic standpoint, the field was presented by \citet{VeblenYoung1910Geometry1,VeblenYoung1910Geometry2}. As to more specific discussions, that might have been directly relevant, we note that Pascal's theorem and especially its different versions are discussed by \citep[95-101]{grassmann_1909}.\footnote{The author is Hermann Ernst Gra{\ss}mann (1857--1922), son of Hermann Günther Gra{\ss}mann (1809--1877), known for his \emph{Ausdehnungslehre}.} His visualizations of the different versions look very similar to Einstein's sketches on AEA 62-787r and in the Scratch Notebook.
\citet{enriques_1915} is noteworthy for its statements about the radical axis and involutions on a line \cite[124-128]{enriques_1915}. Discussion with sketches similar to Einstein's sketches can also be found in older books from the beginning of the 20th century by, for example, \citet[13--14]{emch_1905}, \citet[101]{hatton_1913}, \citet[82]{askwith_1917}, or \citet[131]{dowling_1917}.

In our reconstruction of Einstein's geometric arguments, we will often refer to Coxeter's books on \emph{The Real Projective Plane} from 1949\footnote{We used the second edition \citep{coxeter_1961}.}  and on \emph{Projective Geometry} from 1964.\footnote{We used the second editon \citep{coxeter_1987}.} But, again, the subject matter is treated in many different ways in many different texts.

Finally, we would like to point out that in a little book which treats ``the most famous problems of mathematics'', we found many theorems with visualizations vaguely reminiscent of Einstein's sketches \citep[261--284]{doerrie_1958}. Although the book is from 1958, the first edition was published in 1932. This book may be of interest because of Einstein's predilection for amusing problems in mathematics as addressed in ``Mathematische Mu{\ss}estunden'' by Hermann \citet{schubert_1898}, some of which are found already in the Scratch Notebook, see also \citet{rowe_2011}.

\section{Involution on a line}
\label{sec:inv}

\begin{figure}[htbp]
\centering
\includegraphics[width=.45\textwidth]{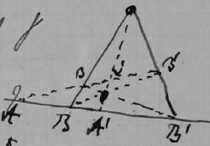}
\hspace{.03\textwidth}
\includegraphics[width=.45\textwidth]{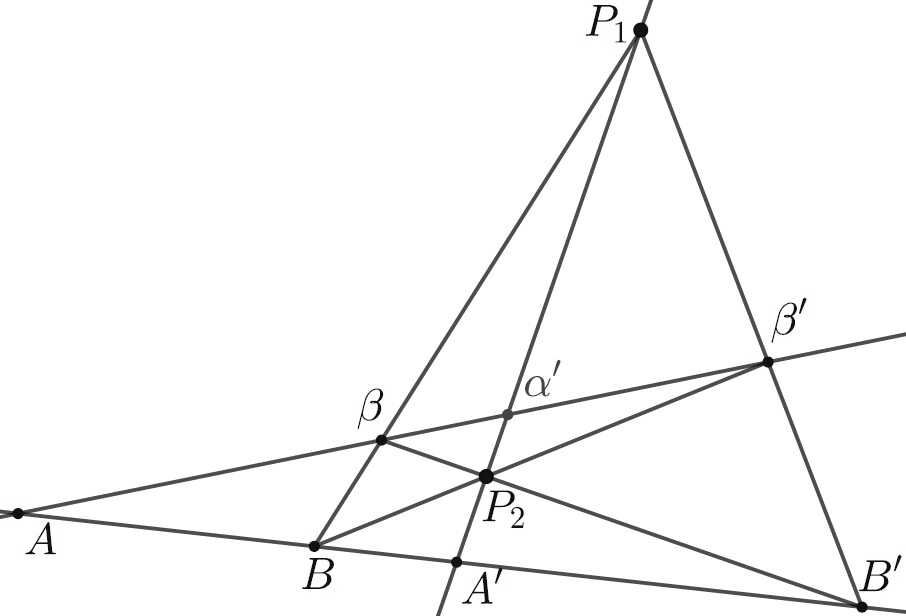}
\caption{Hyperbolic involution with invariant points $A$ and $A'$ that interchanges points $B$ and $B'$. (a) Einstein's first sketch in the Scratch Notebook. Reproduced with permission from the Albert Einstein Archives. \copyright The Hebrew University of Jerusalem, Israel. Digital image photographed by Ardon Bar Hama. (b) Reconstruction of Einstein's sketch.}
\label{fig:notebook_1}
\end{figure}

The sketches in \cref{fig:notebook_1} can be interpreted as a hyperbolic involution on a line that interchanges points $B$ and $B'$, while points $A$ and $A'$ remain invariant. Indeed, by \cite[47]{coxeter_1987}, two invariant points $A$ and $A'$ of an involution are harmonic conjugates with respect to any other pair of the involution. This holds in \cref{fig:notebook_1} by the construction of the complete quadrangle \cite[22]{coxeter_1987}.

Einstein's notation supports this interpretation. He apparently described the involution by the product of two perspectivities. Let us therefore note that Einstein marked points $P_1$ and $P_2$ each by a thick dot. Considering the perspectivity with center $P_1$, point $B$ goes to $\beta$, $A'$ goes to $\alpha'$, and  $B'$ goes to $\beta'$, while the point $A$ remains invariant. Carrying out the second perspectivity with center $P_2$ takes $\beta$  to $B'$, $\alpha'$ back to $A'$, and $\beta'$ to $B$, while $A$ again remains invariant.\footnote{Einstein chose the notation of points $\alpha'$, $\beta$, and $\beta'$ corresponding to points $A$, $B$, and $B'$. The point $\alpha$ is missing in Einstein's sketch, since $A$ is an invariant point.} The product of the two perspectivities, thus, interchanges the points $B$ and $B'$ and lets the points $A$ and $A'$ invariant. Einstein concluded: ``$B$ and $B'$ interchanged'', which he wrote directly underneath the sketch.\footnote{Einstein wrote ``$B$ und $B'$ vert.'' The abbreviation ``vert.'' probably stands for the word ``vertauscht'' (``interchanged'').}

\subsection{Transition to Manuscript Page AEA~124-446}

\begin{figure}[htbp]
\centering
\includegraphics[width=.45\textwidth]{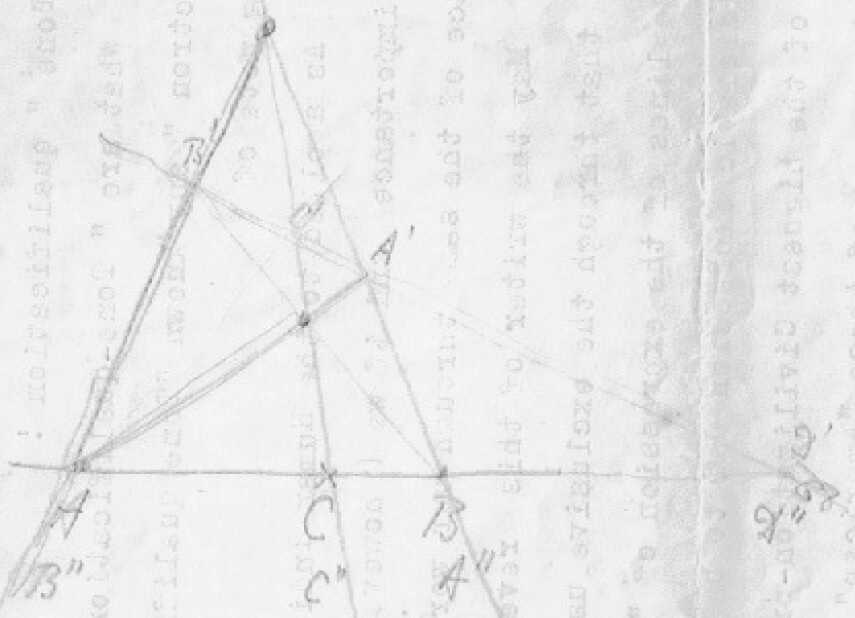}
\hspace{.03\textwidth}
\includegraphics[width=.45\textwidth]{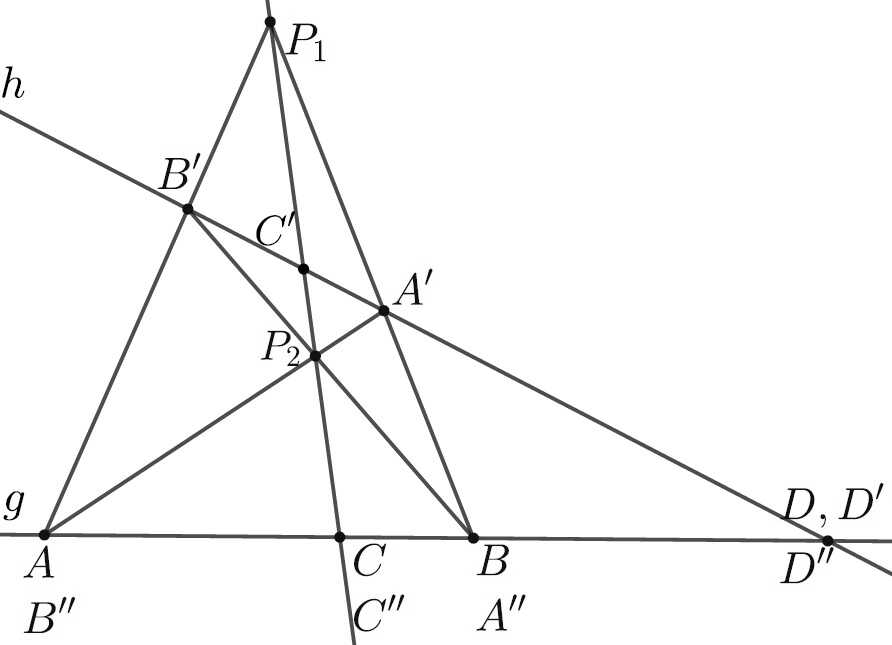}
\caption{Involution with invariant points $C$ and $D$ that interchanges points $A$ and $B$.
(a) Einstein's second sketch on manuscript page AEA~124-446. Reproduced with permission from the Albert Einstein Archives. \copyright The Hebrew University of Jerusalem, Israel. Digital image photographed by Ardon Bar Hama. (b) Reconstruction of Einstein's sketch.}
\label{fig:124446_2}
\end{figure}

Before discussing related sketches in the Princeton manuscripts, we observe that a sketch very similar to the first sketch in the Scratch Notebook can also be found on manuscript page AEA~124-446, see \cref{fig:124446_2}. To show the relation between the two sketches, we start from \cref{fig:notebook_1}, move point $\beta$ up such that point $A$ vanishes at infinity and reappears to meet the base line $g$ on the right hand side of the complete quadrangle.\footnote{The relations between the sketches discussed here and in the following are visualized in the video sequence ``Notebook1.mp4''.} Changing the notation, i.e.\ $B\rightarrow A$, etc.,  leads us to \cref{fig:124446_2}. Einstein's notation suggests that in contrast to \cref{fig:notebook_1} he now carried out the perspectivity with center $P_2$ before the perspectivity with center $P_1$. By the first perspectivity, the point $A$ goes to $A'$, $C$ goes to $C'$, $B$ goes to $B'$ and $D$ remains invariant while Einstein wrote down $D'$ anyhow.\footnote{The notation $D$ is barely readable.} By the second perspectivity with center $P_1$, $A'$ goes to $B$ which is also denoted by $A''$, $B'$ goes to $A$ which is also denoted by $B''$, $C'$ goes back to $C$ which is also denoted by $C''$, and $D'$ remains invariant which is also denoted by $D''$. The product of the two perspectivities thus interchanges the points $A$ and $B$, while the two points $C$ and $D$ remain invariant, in full equivalence to the situation in the Scratch Notebook. The notation here is even more directly indicating the procedure described above.

\subsection{Similar Sketches on Manuscript Pages AEA~62-789 and 62-789r}
\label{subsec:general}

\begin{figure}[htbp]
\centering
\includegraphics[width=.45\textwidth]{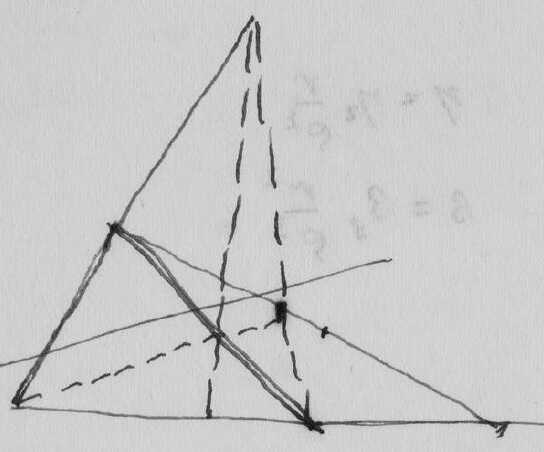}
\hspace{.05\textwidth}
\includegraphics[width=.45\textwidth]{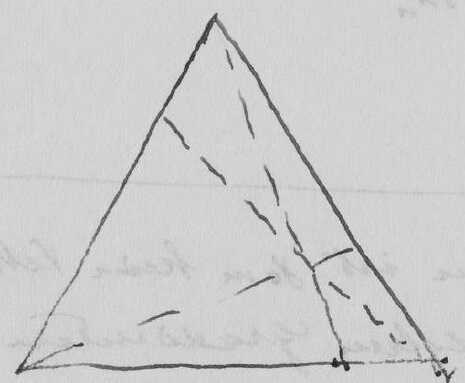}
\caption{Sketches similar to the first sketch in the Scratch Notebook on manuscript pages AEA~62-789 and its reverse side 62-789r. (a) Einstein's first sketch on manuscript page AEA~62-789. It is crossed out; (b) Einstein's fourth sketch on manuscript page AEA~62-789r. Reproduced with permission from the Albert Einstein Archives. \copyright The Hebrew University of Jerusalem, Israel. Digital image photographed by Ardon Bar Hama.}
\label{fig:62789_1}
\end{figure}

Two similar sketches appear on manuscript page AEA~62-789 as well as on its reverse side AEA~62-789r, see \cref{fig:62789_1}.

\begin{figure}[htbp]
\centering
\includegraphics[width=.40\textwidth]{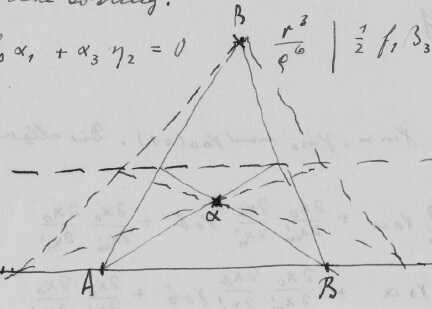}
\hspace{.05\textwidth}
\includegraphics[width=.50\textwidth]{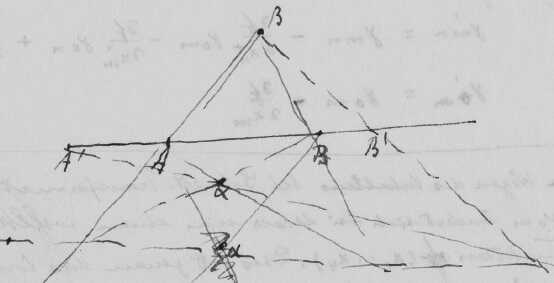}
\caption{Sketches on AEA~62-789, where the fourth harmonic point meets line $g$ in the point at infinity. (a) Einstein's first sketch on manuscript page AEA~62-789r; (b) Einstein's fifth sketch on manuscript page AEA~62-789r. Reproduced with permission from the Albert Einstein Archives. \copyright The Hebrew University of Jerusalem, Israel. Digital image photographed by Ardon Bar Hama.}
\label{fig:62789_2}
\end{figure}

However, Einstein here did not letter any points or lines and it is not as clear as before whether Einstein drew these sketches in the context of hyperbolic involutions. Indeed, both figures are very generic sketches in projective geometry, which allow for many different contexts as, for example, the discussion of the complete quadrangle and the harmonic relation of four points on a line, see \cite[22]{coxeter_1987}.\footnote{In \cref{fig:62789_1}(b), the line for the fourth harmonic point is missing that connects, in terms of \cref{fig:124446_2}, the points $B'$ and $A'$. Next to this sketch, Einstein drew \cref{fig:62789_2}(b), which contains a marked point on the left margin on the horizontal dashed line. This point might belong to the sketch in \cref{fig:62789_1}(b), which then approximately marks the position of the fourth harmonic point.} We will discuss another context in \cref{sec:fifth_sketch}, namely the construction of two invariant points and one pair of an involution.

The very same manuscript pages 62-789 and 62-789r, however, also show sketches that, again, can be related to the figures of the Scratch Notebook, see figure \ref{fig:62789_2}.
Assuming the situation and notation of \cref{fig:124446_2}, we can ask what happens when line $h$ is turned around a bit to become parallel to the base line $g$. Einstein apparently investigated this situation in the first and fifth sketch on AEA~62-789r  (\cref{fig:62789_2}). In this case, the fourth harmonic point $D$ goes to the point at infinity. Since the harmonic relation between the points $A$, $C$, $B$, and $D$ still holds, the point $C$ becomes the midpoint of the segment $AB$, see \cite[119]{coxeter_1961} or \cite[109]{kaplansky_1969}. Einstein drew this situation in \cref{fig:62789_2}(a), where the points $A$ and $B$ are the same as in \cref{fig:124446_2} and the line connecting $P_1$ and $P_2$ yielding the point $C$ is missing.\footnote{Einstein denoted points $P_1$ and $P_2$ in \cref{fig:124446_2}(b) as $\beta$ and $\alpha$ in \cref{fig:62789_2}(a), respectively.}

 Einstein not only drew this situation for the points $A$ and $B$, but also for two points lying further out (see the dashed lines). In \cref{fig:62789_2}(b), he moved the dashed horizontal line underneath the solid horizontal line such that $\alpha$ now lies underneath $A$ and $B$.\footnote{He first accidentally drew  point $\alpha$ incorrectly and crossed it out in \cref{fig:62789_2}(b).} In this situation, he marked the two additional points lying further out by $A'$ and $B'$.\footnote{He did not draw the line connecting $\beta$ and $A'$. As mentioned above, he also marked a point on the left margin on the horizontal dashed line. We argue that this point might belong to \cref{fig:62789_1}(b). However, it also is approximately the intersection between the line $A'\beta$ and the dashed horizontal line. Einstein did not draw \cref{fig:62789_2}(b) accurately, since the distance between $A'$ and $A$ is not equal to the distance between $B$ and $B'$. Furthermore, the two horizontal lines are not parallel. Due to this inaccuracy the line through $\beta$ and $A'$ would actually not pass through Einstein's marked point.}

In addition to figures \ref{fig:62789_1} and \ref{fig:62789_2}, the page contains two more sketches probably written in the context of projective geometry, although we have found no convincing reconstruction for them. One of these sketches has striking similarities to a sketch appearing in Grossmann's lecture notes on ``Projektivische Geometrie: Geometrie der Lage'' held by Fiedler \cite[41]{grossmann_1897}.

\subsection{Another Sketch in the Scratch Notebook}
\label{sec:fifth_sketch}

\begin{figure}[htbp]
\centering
\includegraphics[width=.45\textwidth]{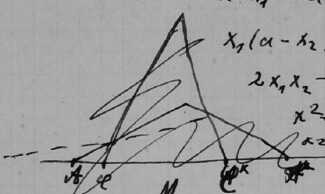}
\hspace{.03\textwidth}
\includegraphics[width=.45\textwidth]{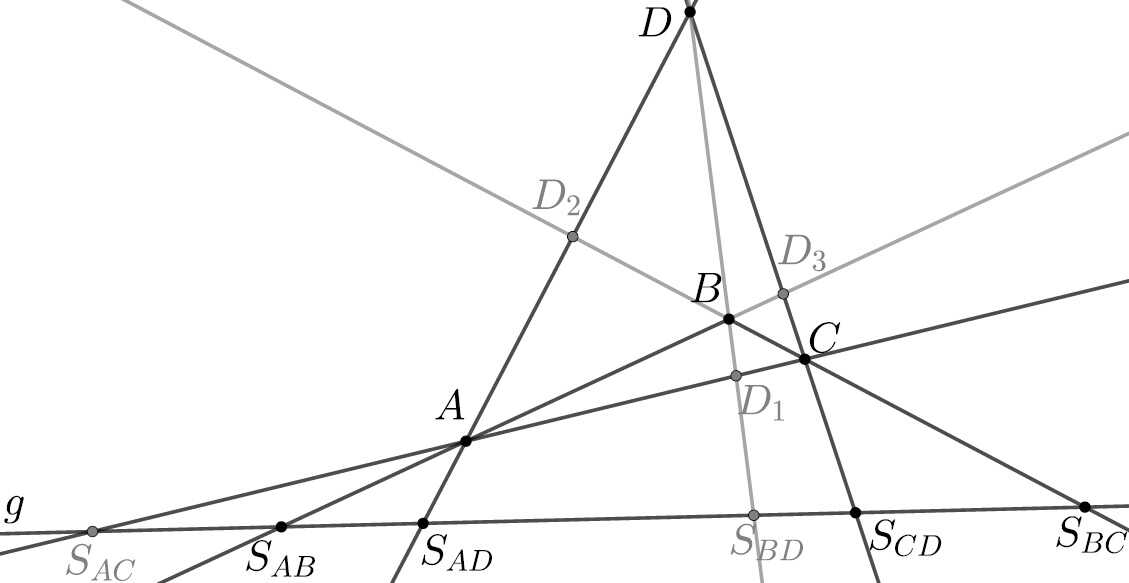}
\caption{The construction of an involution on a line. (a) Einstein's fifth sketch in the Scratch Notebook. It is crossed out. Reproduced with permission from the Albert Einstein Archives. \copyright The Hebrew University of Jerusalem, Israel. Digital image photographed by Ardon Bar Hama. (b) Reconstruction of Einstein's sketch. He did not draw the gray objects.}
\label{fig:notebook_3}
\end{figure}

Let us return to the Scratch Notebook.
Einstein's fifth sketch in the Scratch Notebook is shown in \cref{fig:notebook_3}(a). We will use the notation of our reconstruction in \cref{fig:notebook_3}(b). Considering the complete quadrangle with vertices $A$, $B$, $C$, $D$ and diagonal points $D_1$, $D_2$, $D_3$, opposite sides of the quadrangle\footnote{Opposite sides of the quadrangle are sides that meet each other in a diagonal point and not in a vertex.} meet a line $g$ in pairs of an involution, see \citet[122,123]{enriques_1915} or \citet[269]{doerrie_1958}.\footnote{This theorem can also be found in Grossmann's own lecture notes from 1907 \cite[18.2]{grossmann_1907}. \label{foot:grossmann_1}} This implies that $S_{AB}$ and $S_{CD}$ is a pair of the involution. Einstein denoted them by $A$ and $A^{\ast}$ but overwrote the notation of the second point afterwards. Another pair of the involution are the points $S_{AD}$ and $S_{BC}$, which were denoted as $C$ and $C^{\ast}$ by Einstein. The second point was then overwritten as well. The third pair $S_{AC}$ and $S_{BD}$ was not marked by Einstein. Since Einstein drew the line through $A$ and $C$, the intersection $S_{AC}$ is indicated, while he did not draw the line through $D$ and $B$, which would give the intersection $S_{BD}$. Note that Einstein in his sketch did not use the letter $B$, but did use $A$ and $C$.

It is neither clear why Einstein changed the notation of the two points on the right nor why he crossed out the entire sketch. The initial notation, however, suggests that he again considered the construction of an involution on a line. 

By these considerations, we can get an alternative interpretation of \cref{fig:notebook_1}. Let us consider the complete quadrangle with vertices $\beta$, $P_1$, $\beta'$, $P_2$ and diagonal points $\alpha'$, $B$, and $B'$. In contrast to the situation of figure \ref{fig:notebook_3}, the line $g$ then passes through two diagonal points $B$ and $B'$, which is why we can interpret this sketch as the construction of an involution on the line $g$ with the pair $A$ and $A'$ and the invariant points $B$ and $B'$.  This interpretation does not explain Einstein's comment that the points $B$ and $B'$ are interchanged nor does it explain the notation chosen by Einstein nor why he bolded points $P_1$ and $P_2$. However, it is possible that Einstein considered it this way in the process of drawing \cref{fig:notebook_3}(a). By moving the point $P_2$ in \cref{fig:notebook_1}(b) slightly to the upper right, we get \cref{fig:notebook_3}.\footnote{The relations between the sketches discussed here and in the following are visualized in the video sequence ``Notebook3.mp4''.} As we will see in the next section, it stands to reason that Einstein considered such a transition on the manuscript page AEA~62-789 as well.

\subsection{Similarities to Manuscript Page AEA~62-789}

\begin{figure}[htbp]
\centering
\includegraphics[width=.45\textwidth]{files/62789_1.jpg}
\hspace{.05\textwidth}
\includegraphics[width=.45\textwidth]{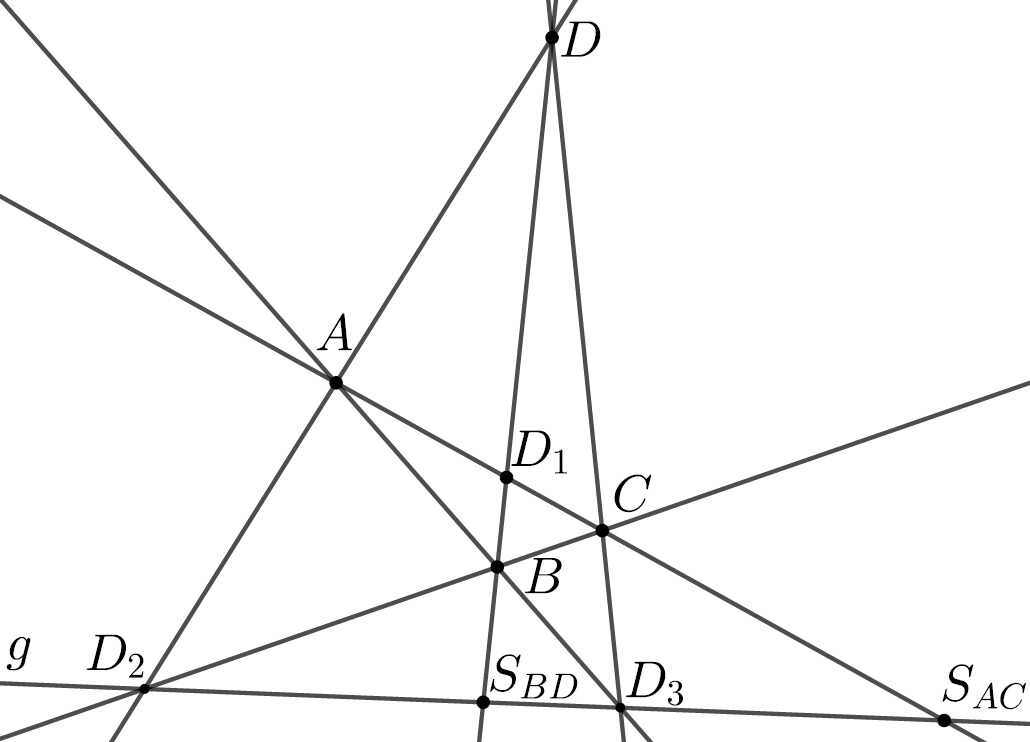}
\caption{The first sketch on manuscript page AEA~62-789, interpreted as the construction of a pair and two invariant points of an involution on a line. (a) Einstein's first sketch on manuscript page AEA~62-789. It is crossed out. Reproduced with permission from the Albert Einstein Archives. \copyright The Hebrew University of Jerusalem, Israel. Digital image photographed by Ardon Bar Hama. (b) Reconstruction of Einstein's sketch.}
\label{fig:62789_3}
\end{figure}

\begin{figure}[htbp]
\centering
\includegraphics[width=.4\textwidth]{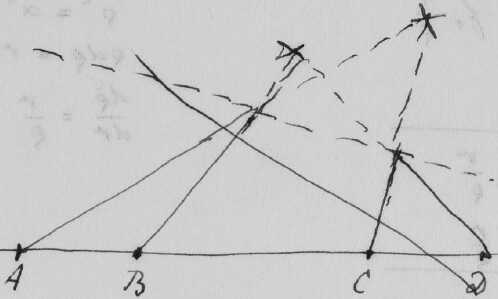}
\hspace{.03\textwidth}
\includegraphics[width=.4\textwidth]{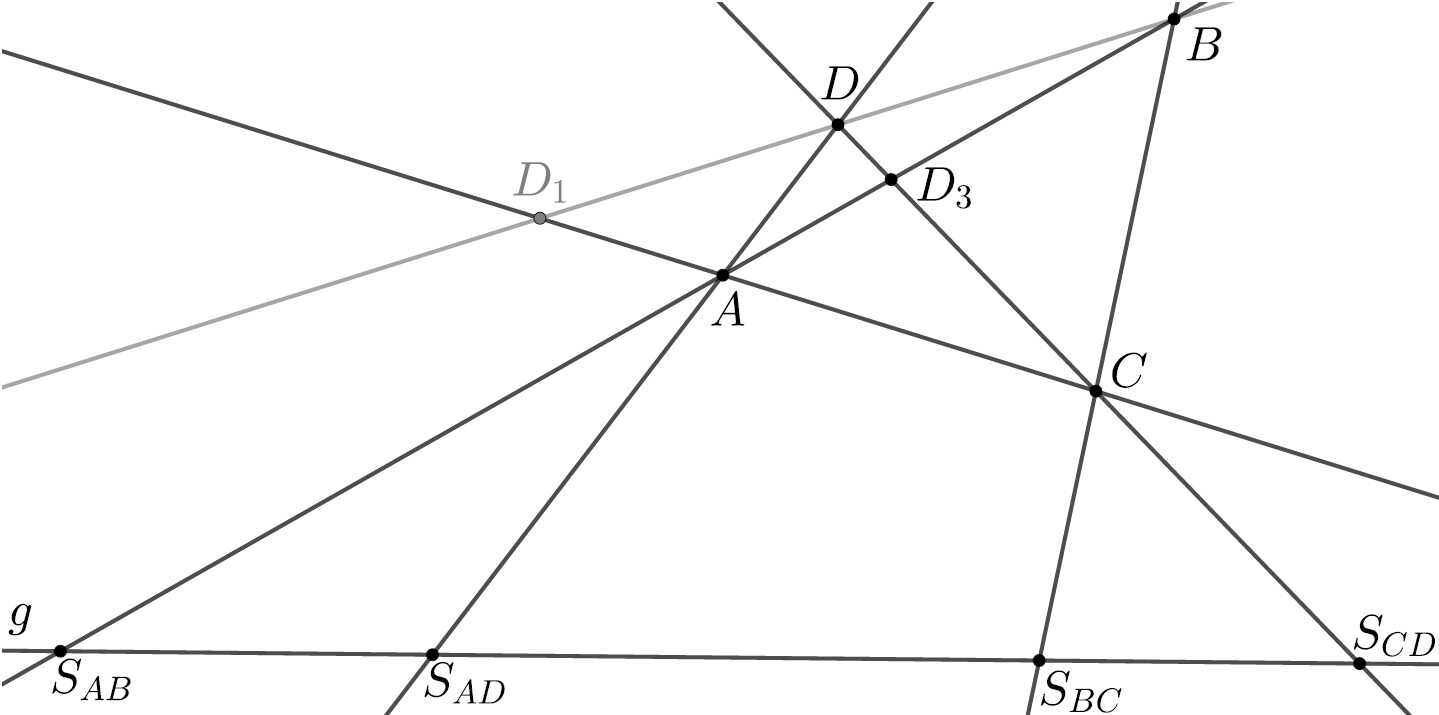}
\caption{The second sketch on manuscript page AEA~62-789, interpreted as the construction two pairs of an involution on a line, see also \cref{fig:notebook_3}. (a) Einstein's second sketch on manuscript page AEA~62-789. It is crossed out. Reproduced with permission from the Albert Einstein Archives. \copyright The Hebrew University of Jerusalem, Israel. Digital image photographed by Ardon Bar Hama. (b) Reconstruction of Einstein's sketch. He did not draw the gray lines.}
\label{fig:62789_4}
\end{figure}

If we turn over page 62-789r to look at 62-789, we find two sketches next to each other at the top of manuscript page. These again can be connected to the sketches in figures \ref{fig:notebook_1}(a) and \ref{fig:notebook_3}(a) in the Scratch Notebook. They are shown in \cref{fig:62789_4,fig:62789_3}.\footnote{We already know \cref{fig:62789_4}(a) from \cref{fig:62789_1}(a).}

By introducing notation as in \cref{fig:62789_3}(b) and considering the complete quadrangle with vertices $A$, $B$, $C$, $D$ and diagonal points $D_1$, $D_2$, $D_3$, the points $D_2$ and $D_3$ can be interpreted as invariant points of an involution on the line $g$, while points $S_{BD}$ and $S_{AC}$ are corresponding points. 

If we move the point $B$ to the upper right, we directly get \cref{fig:62789_4}, where the line $g$ does not pass through any diagonal point anymore. Thus, the points $S_{AB}$, $S_{CD}$ and $S_{AD}$, $S_{BC}$ are corresponding points, respectively. Equivalently to \cref{fig:notebook_3}, Einstein drew one more line whose intersection with $g$ determines one point of the third pair (line $AC$).

We conclude that all sketches from figures \ref{fig:notebook_1}--\ref{fig:62789_4} can be interpreted in some way or other as constructions involving involutions on a line.

We also note that just as we find figures \ref{fig:notebook_1}(a) and \ref{fig:notebook_3}(a) next to each other on a double page of the Scratch Notebook, two very similar sketches appear right next to each other at the top of manuscript page AEA~62-789.

\section{Pascal's Theorem and Involution on a Conic}
\label{sec:pas}

Surprisingly, we find a similar correspondence between sketches in the Scratch Notebook and manuscript page 62-787. These sketches deal with projective geometry on conics.

\begin{figure}[htbp]
\centering
\includegraphics[width=.45\textwidth]{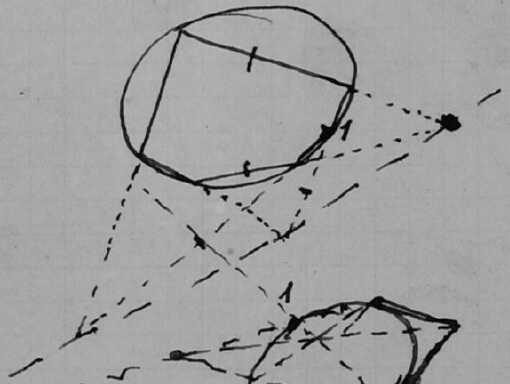}
\hspace{.03\textwidth}
\includegraphics[width=.45\textwidth]{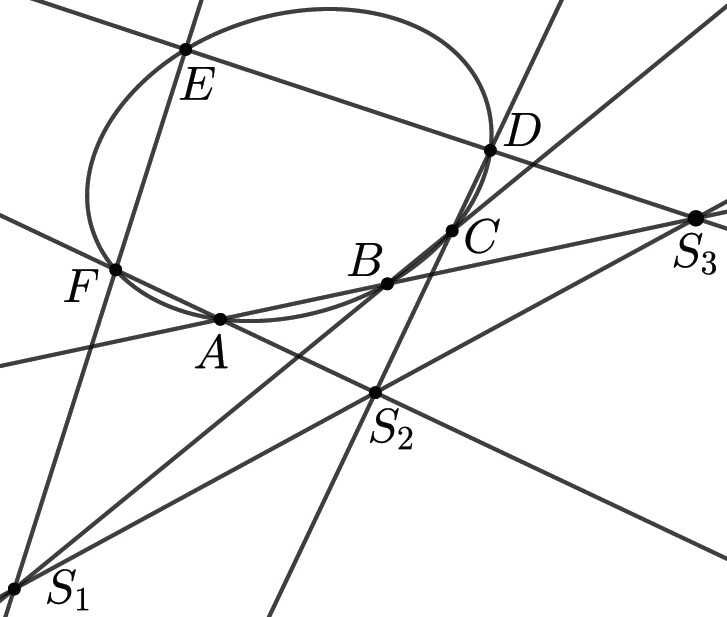}
\caption{Pascal's theorem. Opposite sides of a hexagon inscribed in a conic meet in three collinear points. (a) Einstein's second sketch in the Scratch Notebook. Reproduced with permission from the Albert Einstein Archives. \copyright The Hebrew University of Jerusalem, Israel. Digital image photographed by Ardon Bar Hama. (b) Reconstruction of Einstein's  sketch.}
\label{fig:notebook_4}
\end{figure}

Einstein's Scratch Notebook shows a second sketch, which is shown in \cref{fig:notebook_4} together with our reconstruction. As the only numbered point he denoted point $C$ by $1$. He also drew point $S_3$ as a thick dot and marked lines $ED$ and $AB$. The sketch visualizes Pascal's theorem which states that opposite sides of a hexagon $ABCDEF$ inscribed in a conic\footnote{The opposite sides of a hexagon $ABCDEF$ are $BC$ and $EF$, $CD$ and $AF$ as well as $AB$ and $DE$.} meet each other in three collinear points $S_1$, $S_2$, and $S_3$ \cite[103]{coxeter_1961}. The line passing through these points is called \textit{Pascal line}.

\begin{figure}[htbp]
\centering
\includegraphics[width=.45\textwidth]{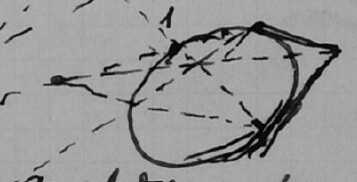}
\hspace{.03\textwidth}
\includegraphics[width=.45\textwidth]{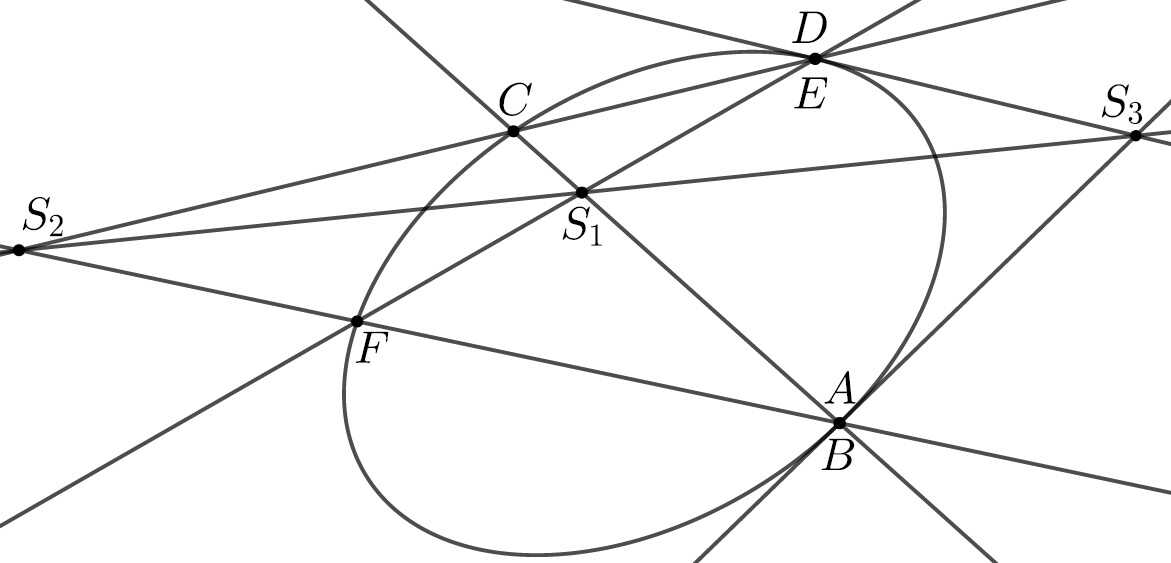}
\caption{Pascal's theorem for a quadrangle. (a) Einstein's third sketch in the Scratch Notebook. Reproduced with permission from the Albert Einstein Archives. \copyright The Hebrew University of Jerusalem, Israel. Digital image photographed by Ardon Bar Hama. (b) Reconstruction of Einstein's third sketch in the Scratch Notebook.}
\label{fig:notebook_5}
\end{figure}

The sketch just underneath shows a quadrangle inscribed in a conic, see \cref{fig:notebook_5}. Again, by our reconstruction, point $C$ was numbered $1$ by Einstein. By \cite[99]{grassmann_1909}, in this situation the tangents in $A$ and $D$ as well as the opposite sides $AF$, $DC$ and $AC$, $DF$ meet each other in three collinear points $S_1$, $S_2$, $S_3$.

Steiner's theorem enables us to define projectivities on conics \cite[105]{coxeter_1961}, which again allows us to interpret Einstein's sketches as investigations of projectivities and involutions. \Cref{fig:notebook_4}(b) can be interpreted as the projectivity $AEC \barwedge DBF$. In this particular case, the Pascal line becomes the axis of the projectivity.\footnote{Knowing the axis of the projectivity, any other pair of the projectivity can be constructed \cite[106]{coxeter_1961}. We generally define the axis of the projectivity $ABC \barwedge A'B'C'$ as the Pascal line of the hexagon $AB'CA'BC'$ inscribed in a conic.} 

When moving toward each other points $A$ and $B$ as well as $E$ and $D$, respectively,  the secants $AB$ and $ED$ in \cref{fig:notebook_4} become tangents which meet in $S_3$. This procedure was indicated by Einstein by marking these two lines as well as point $S_3$. When furthermore moving point $C$ along the conic such that it comes to lie between points $F$ and $E=D$, we obtain Einstein's sketch in \cref{fig:notebook_5}. Moving point $C$ into this position was also indicated by Einstein since it is the only point numbered in both \cref{fig:notebook_4}(a) and \cref{fig:notebook_5}(a).\footnote{The transition is also visualized in the video sequence ``Notebook2.mp4''.}

The projectivity $AEC \barwedge DBF$ from \cref{fig:notebook_4} then becomes $ADC \barwedge DAF$. This means that the projectivity interchanges points $A$ and $D$ and thus is an involution \cite[45]{coxeter_1987}. By knowing another pair $C$, $F$ of the involution, it is fully determined \cite[45]{coxeter_1987}. As we see in \cref{fig:notebook_5}, the Pascal line furthermore crosses the conic in two points which are the invariant points of the involution,\footnote{This becomes clear by the construction of corresponding points using the Pascal line \cite[106]{coxeter_1961}. A hyperbolic involution has two invariant points, while an elliptic involution has no real invariant points. A so called parabolic involution has only one invariant point.} which therefore is a hyperbolic one. 

Beneath the third sketch, Einstein wrote: ``At the same time construction of the center.''\footnote{``Zugleich Konstruktion des Zentrums.''} Any secant through the center of the involution determines a pair of the involution as the two intersections with the conic \cite[108-109]{coxeter_1961}.\footnote{We note that drawing two tangents through the center of the involution thus determines the two invariant points, which are also the intersections between Pascal line and conic. In the case that the center of the involution lies inside the conic, the Pascal line is an exterior line and the involution is elliptic without any real invariant points. In \cref{fig:notebook_5}, the center of the involution is an exterior point.} Since \cref{fig:notebook_5} can be interpreted as the involution $ADC \barwedge DAF$,  the intersection of the lines $AD$ and $CF$ determines the center of the involution. This might be the idea of Einstein's additional comment, since he did not draw these lines.

\subsection{Similarities to Manuscript Page AEA~62-787r}
\label{subsec:62787r}

The manuscript page AEA~62-787r contains sketches that are astonishingly similar to the second and third sketch in the Scratch Notebook that we have just discussed. Moreover, this page illustrates the ideas from the previous section by a more precise notation as well as by additional sketches.

\begin{figure}[htbp]

\centering
\includegraphics[width=.30\textwidth]{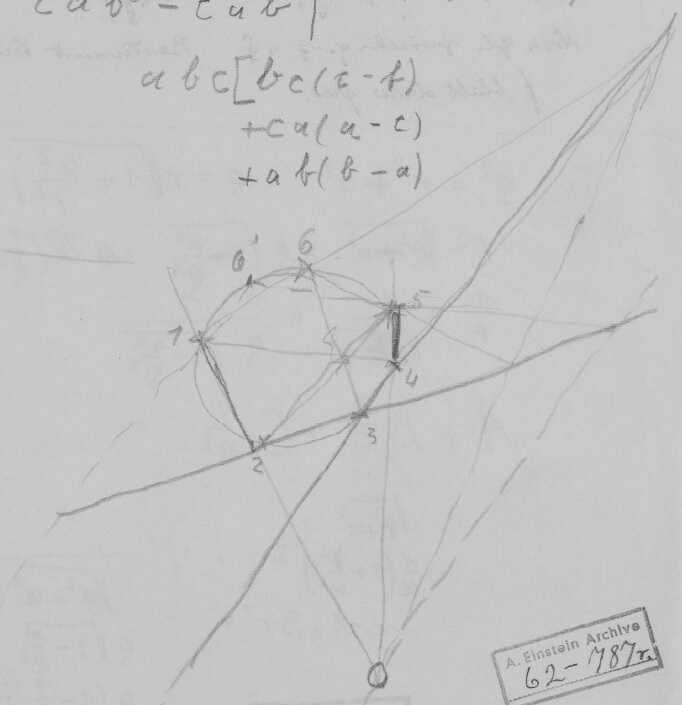}
\hspace{.03\textwidth}
\includegraphics[width=.30\textwidth]{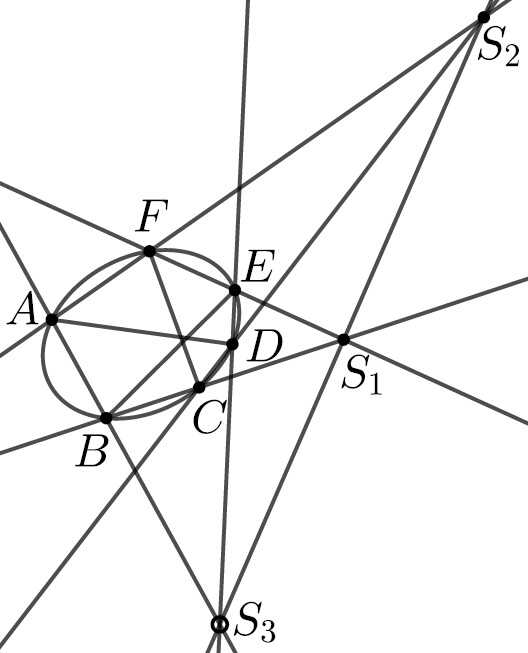}
\hspace{.03\textwidth}
\includegraphics[width=.30\textwidth]{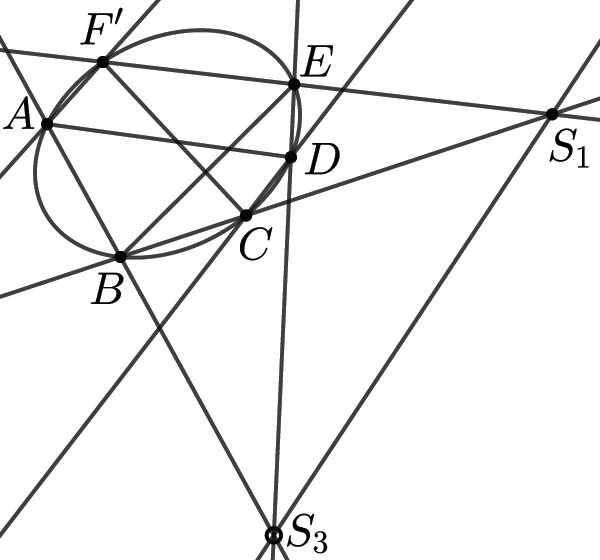}
\caption{Pascal's theorem. (a) Einstein's sketch on the right hand side of manuscript page AEA~62-787r. Reproduced with permission from the Albert Einstein Archives. \copyright The Hebrew University of Jerusalem, Israel. Digital image photographed by Ardon Bar Hama. (b) Reconstruction of Einstein's sketch ignoring point $6'$; (c) Reconstruction of Einstein's sketch ignoring point $6$.}
\label{fig:62787r_1}
\end{figure}

Let us start with the sketch on the right hand side of the manuscript page, see \cref{fig:62787r_1}(a). It shows Pascal's theorem, just as in figure \ref{fig:notebook_4}, but Einstein here combined two different situation into this sketch. The first situation is shown in \cref{fig:62787r_1}(b). Einstein numbered  points on the conic from $1$ to $6$, while we denoted them by $A$ to $F$. The sketch therefore is an identical repeat of illustrating Pascal's theorem as in \cref{fig:notebook_4}. Einstein then replaced point $F$ by $F'$ such that the opposite lines $AF'$ and $CD$ become parallel.\footnote{The transition from the point $F$ to $F'$ is visualized in the video sequence ``62787rPoint6.mp4''.} They then meet in the point at infinity $S_2$. By Pascal's theorem, the Pascal line which connects $S_1$ and $S_3$ needs to pass through the point at infinity $S_2$ as well, which is why it becomes parallel to these opposite sides. 

By these considerations, we can explain most of the lines drawn by Einstein. However, Einstein also drew the line segments $AD$, $BE$, and $CF$ inside the conic.\footnote{Although it appears in \cref{fig:62787r_1}(a) and \cref{fig:62787r_1}(b) as if the three lines meet each other in one common point, this is in general not the case.} These segments indicate that Einstein considered the projectivity $AEC \barwedge DBF$. In this case, he connected the corresponding points by a line segment. Considering this projectivity, the axis of the projectivity becomes the Pascal line, which is drawn by Einstein.

As in the Scratch Notebook, Einstein marked the point $S_3$. Moreover, he also bolded lines $ED$ and $AB$. In order to get \cref{fig:62787r_1}(b) starting from \cref{fig:notebook_4}(b), we only need to move some points along the conic without the need of changing the orientation.\footnote{The transitions between the sketches in the notebook and on AEA~62-787r are shown in the video sequence ``NotebookToMS.mp4''.}

\begin{figure}[htbp]
\centering
\includegraphics[width=.34\textwidth]{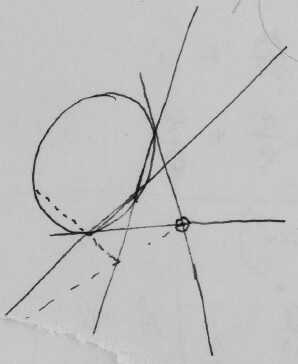}
\hspace{.03\textwidth}
\includegraphics[width=.34\textwidth]{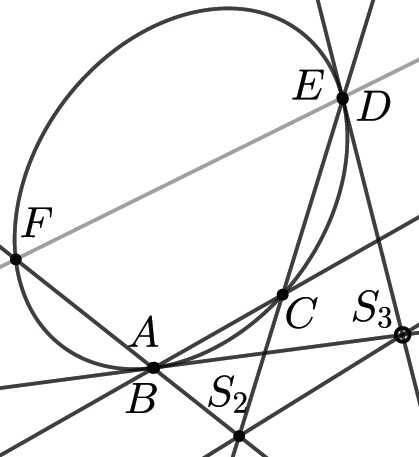}
\caption{Elliptic involution on a conic. (a) Einstein's sketch in the bottom left corner of manuscript page AEA~62-787r, which was not drawn accurately. Reproduced with permission from the Albert Einstein Archives. \copyright The Hebrew University of Jerusalem, Israel. Digital image photographed by Ardon Bar Hama. (b) Reconstruction of Einstein's sketch. He did not draw the gray line $EF$.}
\label{fig:62787r_2}
\end{figure}

As in the discussion of the second and third sketch in the Scratch Notebook, we can now let the points $A$, $B$ and $E$, $D$ approach each other. These two lines then become tangents meeting in the marked point $S_3$. By slightly rearranging the points on the conic, we then get the situation in \cref{fig:62787r_2}, which is another sketch on manuscript page 62-787r. We see that Einstein also marked point $S_3$ in this sketch indicating the transition just described.\footnote{The transitions between the sketches on manuscript page AEA~62-787r are visualized in the video sequence ``MS62787r.mp4''.}

In contrast to the discussion of the Scratch Notebook (figures \ref{fig:notebook_4} and \ref{fig:notebook_5}), we have not moved point $C$ yet. As a result, we do not get a hyperbolic involution, but an elliptic involution $ADC \barwedge DAF$. The Pascal line that goes through $S_2$ and $S_3$ is an exterior line without intersections with the conic such that no invariant point exists.\footnote{The center of the involution lies inside the conic as the intersection point of the lines $CF$ and $AD$, but was not drawn by Einstein.} We note that Einstein's sketch is not accurate as the intersection $S_1$ of $EF$ and $BC$ would meet on the right side of the conic in Einstein's sketch, while the Pascal line meets the line $BC$ rather on the left side of the conic (the piece that had been torn off).\footnote{In our reconstruction in \cref{fig:62787r_2}(b), the lines $EF$, $BC$, and $S_2S_3$ are almost parallel.}

\begin{figure}[htbp]
\centering
\includegraphics[width=.34\textwidth]{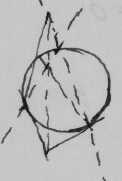}
\hspace{.03\textwidth}
\includegraphics[width=.34\textwidth]{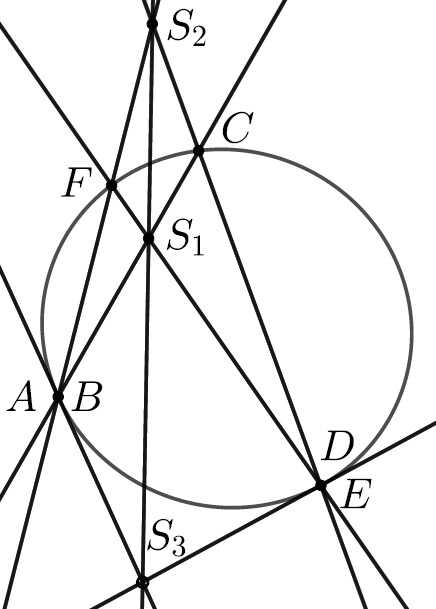}
\caption{Hyperbolic involution on a conic. (a) Einstein's top left sketch on manuscript page AEA~62-787r. Reproduced with permission from the Albert Einstein Archives. \copyright The Hebrew University of Jerusalem, Israel. Digital image photographed by Ardon Bar Hama. (b) Reconstruction of Einstein's sketch.}
\label{fig:62787r_3}
\end{figure}

We can now proceed by moving point $C$ such that it comes to lie between $E=D$ and $F$. By changing the orientation, we then get the situation in \cref{fig:62787r_3}, also found on 62-787r. This sketch can be interpreted as a hyperbolic involution $ADC \barwedge DAF$, since the Pascal line intersects the conic in two points.\footnote{The center of the involution is an exterior point, which was not drawn by Einstein.} We only need to rotate this sketch by $90^{\rm o}$ in order to get \cref{fig:notebook_5} from the Scratch Notebook.

\begin{figure}[htbp]
\centering
\includegraphics[width=.45\textwidth]{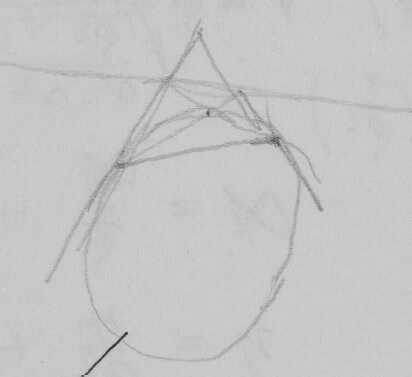}
\hspace{.03\textwidth}
\includegraphics[width=.45\textwidth]{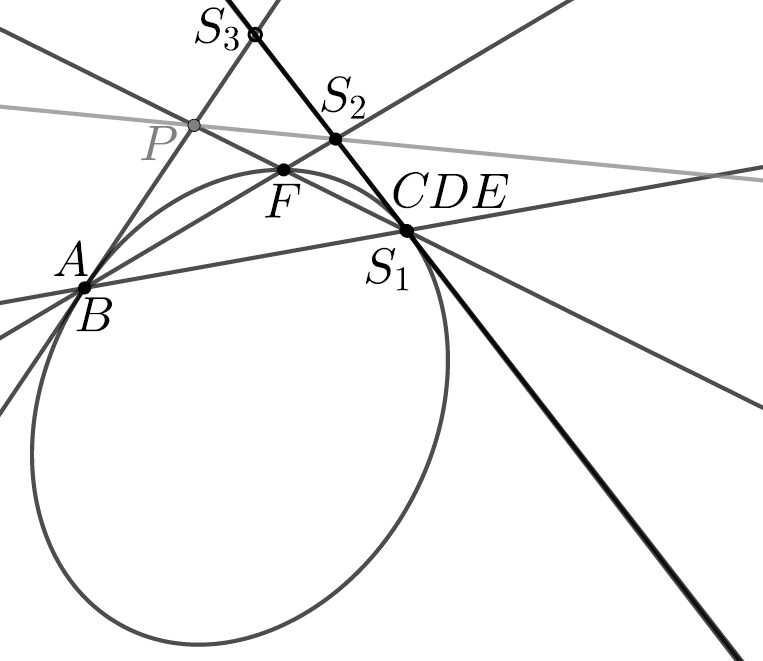}
\caption{Parabolic involution on a conic. (a) Einstein's sketch in the middle of manuscript page AEA~62-787r. Reproduced with permission from the Albert Einstein Archives. \copyright The Hebrew University of Jerusalem, Israel. Digital image photographed by Ardon Bar Hama. (b) Reconstruction of Einstein's sketch. The gray objects were drawn by Einstein, but have no meaning in terms of involutions.}
\label{fig:62787r_4}
\end{figure}

On manuscript page AEA~62-787r, Einstein even went one step further and drew a triangle inscribed in the conic, see \cref{fig:62787r_4}. By \cite[100--101]{grassmann_1909}, the sides of the triangle meet the tangents through the respective opposite vertices in three collinear points. For instance, the tangent in $A$ meets the side $FC$ in $P$, while the tangent in $D$ meets the side $AF$ in $S_2$. A very similar sketch can be found in Grossmann's lecture notes on ``Geometrische Theorie der Invarianten I'' held by Carl Friedrich Geiser and attended by Einstein in 1898 \cite[49]{grossmann_1898b}, \cite[366]{cpae1}.\footnote{A sketch very similar to \cref{fig:62787r_4} can also be found in Grossmann's own lecture notes from 1907 \cite[20.1]{grossmann_1907}. \label{foot:grossmann_2}}

We get to this figure by moving the points $D=E$ and $C$ towards each other in \cref{fig:62787r_3}. In this case, the points $C$, $D$, $E$, and $S_1$ fall together, see \cref{fig:62787r_4}. Thus, the Pascal line passing through the points $S_i$ becomes a tangent, leading to the situation where only one invariant point exists. Since the center of the involution as intersection of the sides $CF$ and $AD$ falls together also with these points, every secant through the center of the involution passes through the point $C=D=E$. Thus, every point on the conic is a corresponding point to $C=D=E$ (parabolic involution). In the strict sense, this is not an involution anymore since it is not a one-to-one-correspondence \cite[52]{coxeter_1961}. Nevertheless, we could describe the involution by $ACC \barwedge CAF$.

\section{Projective Geometry Calculation in the Scratch Notebook}
\label{sec:cal}

\begin{figure}[htbp]
\centering
\includegraphics[width=.45\textwidth]{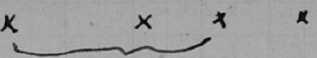}
\hspace{.05\textwidth}
\includegraphics[width=.45\textwidth]{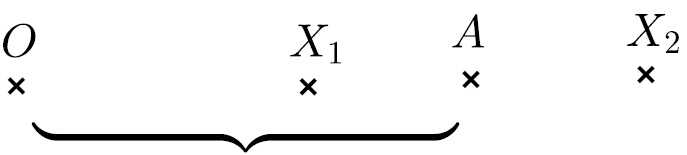}
\caption{Einstein's sketch probably related to the calculation on hyperbolic involutions. (a) Einstein's fourth sketch in the Scratch Notebook. Reproduced with permission from the Albert Einstein Archives. \copyright The Hebrew University of Jerusalem, Israel. Digital image photographed by Ardon Bar Hama. (b) Einstein's sketch with our suggested notation.}
\label{fig:62787r_6}
\end{figure}

The fourth sketch of the double page in the Scratch Notebook, i.e the sketch at the top of the right hand side,  is shown in \cref{fig:62787r_6}(a). Beneath we find a short calculation and the sketch from \cref{fig:notebook_3}(a) that we interpreted as the construction of pairs of an involution on a line. It is a priori not clear whether \cref{fig:62787r_6}(a) is related at all to Einstein's calculation, since it is a very generic sketch. But as we will show, imposing the notation from \cref{fig:62787r_6}(b) enables us to  connect the sketch directly with the calculation, while at the same time the drawn brace gets a meaning as correlating the two invariant points of an involution.

We assume that the points in \cref{fig:62787r_6}(b) lie on one line and the coordinates of the points $O$, $X_1$, $A$, and $X_2$ be $0$, $x_1$, $a$, and $x_2$, respectively.\footnote{In the projective plane, for instance, we could choose $O = (0 : 0 : 1)$, $X_1 = (x_1 : 0 : 1)$, $A = (a : 0 : 1)$, and $X_2 = (x_2 : 0 : 1)$, all lying on the line $x_2 = 0$. \label{foot:projcoord}} A pair $x_1$, $x_2$ of an involution on this line can be described analytically by
\begin{align}
\label{eq:involution}
r x_1 x_2 + s ( x_1 + x_2 ) + t = 0,
\end{align}
with real numbers $r$, $s$, and $t$ which completely describe the involution, see \cite[118,151]{coxeter_1987} or \cite[36]{faulkner_1960}. Imposing the assumption that both $O$ and $A$ are invariant points, we get $t=0$ as well as $s = -a$ when setting $r=2$. Interpreting $X_1$ and $X_2$ as a pair of this involution, the relation
\begin{align}
\label{eq:involution1}
2x_1 x_2 - a ( x_1 + x_2 ) = 0
\end{align}
holds, which can be recast to read
\begin{align}
\label{eq:involution2}
\frac{x_1}{a - x_1} : \frac{x_2}{a-x_2} = -1.
\end{align}
Considering the cross ratio
\begin{align}
( OA, X_1 X_2 ) = \frac{x_2 - a}{x_2} : \frac{x_1 - a}{x_1}
\end{align}
of the points $O$, $A$, $X_1$, and $X_2$ \cite[107]{kaplansky_1969}, \cref{eq:involution2} implies that $O$ and $A$ are harmonic conjugates to $X_1$ and $X_2$. This result is no surprise since the two invariant points of a hyperbolic involution are harmonic conjugates to any other pair of the involution \cite[47]{coxeter_1987}.

Einstein worked the calculation the other way around. Starting from \cref{eq:involution2} he derived \cref{eq:involution1}. He then implicitly set $x = x_1 = x_2$ and got 
\begin{align}
x^2-ax=0.
\end{align}
He concluded
\begin{align}
x=0 \; \text{and} \; x=a.
\end{align}
We summarize that Einstein started with the harmonic relation of four points with the coordinates $0$, $a$, $x_1$, and $x_2$ and recast this equation in order to get the equation of a hyperbolic involution. He then determined the two invariant points by setting $x=x_1=x_2$. He finished his calculation with a word that reads ``Doppelv.'' or ``Doppelp.'', which could either stand for \textit{cross ratio}\footnote{From the German term \textit{Doppelverhältnis}.} or \textit{double point}\footnote{From the German term \textit{Doppelpunkt}. \label{foot:doppelpunkt}}. The latter describes invariant points as in \cite[111]{grassmann_1909}, \cite[68]{enriques_1915} or \cite[32]{coxeter_1961}. As we know from Grossmann's lecture notes, Fiedler used this term in his lecture attended by Einstein as well \cite[71]{grossmann_1897}.\footnote{We will refer to this passage again when discussing the real and imaginary cases in \cref{subsec:62785}.} Both cases fit to the context presented here as Einstein started with the cross ratio at the beginning and determined the invariant points of the involution at the end.

\subsection{Similarities to the Manuscript Page AEA~62-785}
\label{subsec:62785}

On manuscript page AEA~62-785, Einstein made calculations on involutions using \cref{eq:involution} as well. Just as we found the construction of pairs of an involution right next to Einstein's calculation in the Scratch Notebook (see \cref{fig:notebook_3}), Einstein constructed pairs of an involution on AEA~62-785, too. As we will see, he also constructed invariant pairs explicitly, analogous to the interpretation of \cref{fig:notebook_1} in \cref{sec:fifth_sketch}. 

Einstein started with the equation of involution\footnote{Cp.\ \cref{eq:involution}.}
\begin{align}
\label{eq:62785inv_1}
a x y + b ( x + y ) + c = 0
\end{align}
with real numbers $a \neq 0$, $b$, and $c$.\footnote{This time, we also allow the coordinates to be infinite corresponding to the point at infinity. For instance, such a point could have the homogeneous coordinates $(1 : 0 : 0)$ lying on the same line $x_2 = 0$ as the points from \cref{foot:projcoord}, while $x_3 = 0$ could be the line at infinity.} Einstein then investigated the corresponding point of the point at infinity. Letting $x$ be the point at infinity, we have
\begin{align}
\label{eq:62785inv_1_2}
y = - \frac{b}{a}
\end{align}
for the corresponding point $y$. We can either see this by introducing homogeneous coordinates in the projective plane or, as Einstein apparently did, by setting $x=\infty$ after dividing \cref{eq:62785inv_1} by $x$.\footnote{The point $( x_1 : x_2 : x_3)$ in the projective plane has affine coordinates $( x_1 / x_2, x_2 / x_3 )$ for $x_3 \neq 0$. In the case $x_3 = 0$, Einstein apparently considered the affine coordinates to become infinite.} By \cite[121,122]{coxeter_1961}, this point $y$ is called the \textit{center of the involution}. Einstein indicated that this point should be the origin.\footnote{Einstein wrote: ``Reeller Punkt. Als Anfang genommen.''} Setting $y = 0$ yields $b=0$, as written down by Einstein. 

For any pair $x_1$, $y_1$ of the involution, it thus is
\begin{align}
x_1 y_1 + \widetilde{c} = 0
\end{align}
with $\widetilde{c} = \sfrac{c}{a}$. Einstein simply wrote down
\begin{align}
\label{eq:62785inv_2}
xy + c = 0,
\end{align}
implicitly changing the meaning of $c$. By the conditions imposed by Einstein, \cref{eq:62785inv_2} describes the \textit{constant of the involution}, saying that the product $x \cdot y$ of any pair $x$, $y$ is constant. In other words: The distance between the center of the involution and $x$ times the distance between the center of the involution and $y$ is constant \cite[122]{coxeter_1961}.\footnote{Einstein chose the center of the involution to be the origin of the line.} 

\begin{figure}[htbp]
\centering
\includegraphics[width=.45\textwidth]{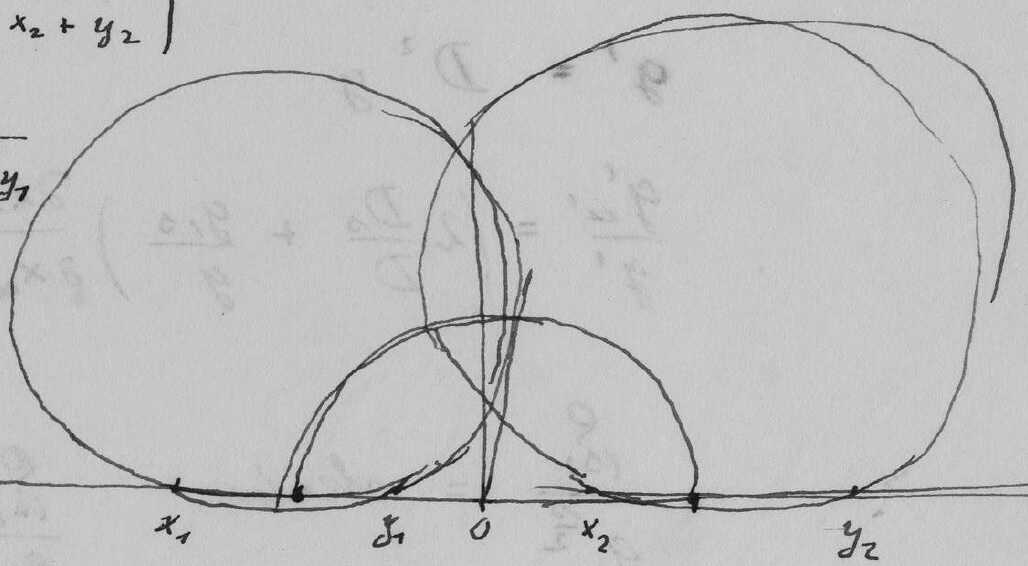}
\hspace{.05\textwidth}
\includegraphics[width=.45\textwidth]{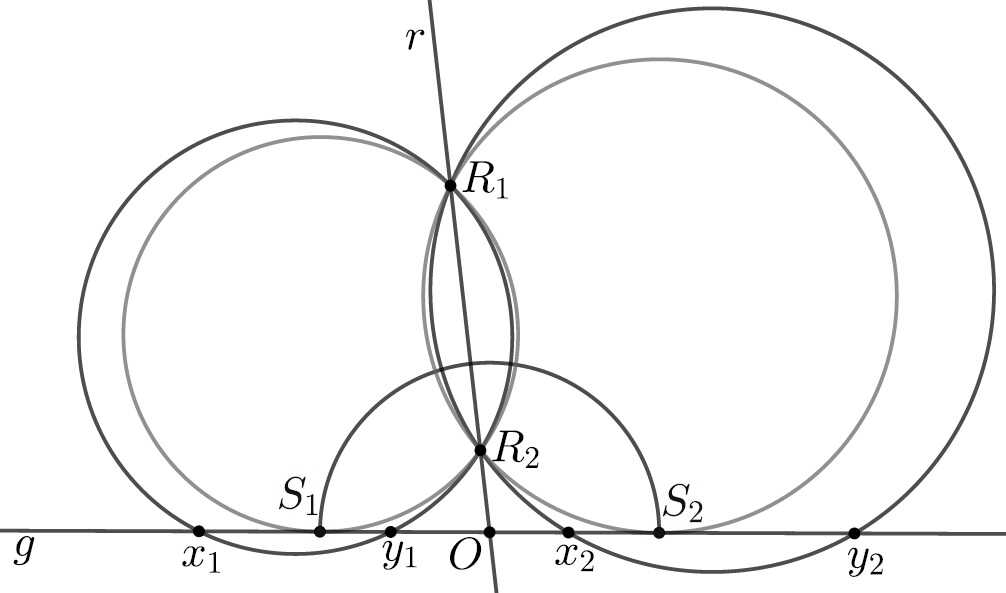}
\caption{Construction of pairs $x_1$, $y_1$ and $x_2$, $y_2$ of a hyperbolic involution on a line as well as construction of the invariant points. (a) Einstein's first sketch on AEA~62-785r. Reproduced with permission from the Albert Einstein Archives. \copyright The Hebrew University of Jerusalem, Israel. Digital image photographed by Ardon Bar Hama. (b) Reconstruction of Einstein's sketch. Einstein did not draw the gray circles.}
\label{fig:785_1}
\end{figure}

\begin{figure}[htbp]
\centering
\centering
\includegraphics[width=.30\textwidth]{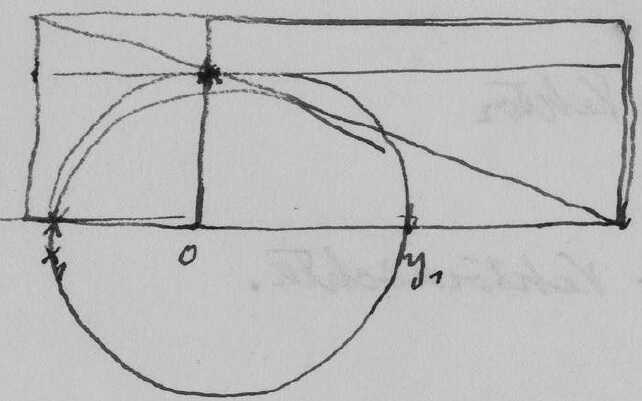}
\hspace{.03\textwidth}
\includegraphics[width=.30\textwidth]{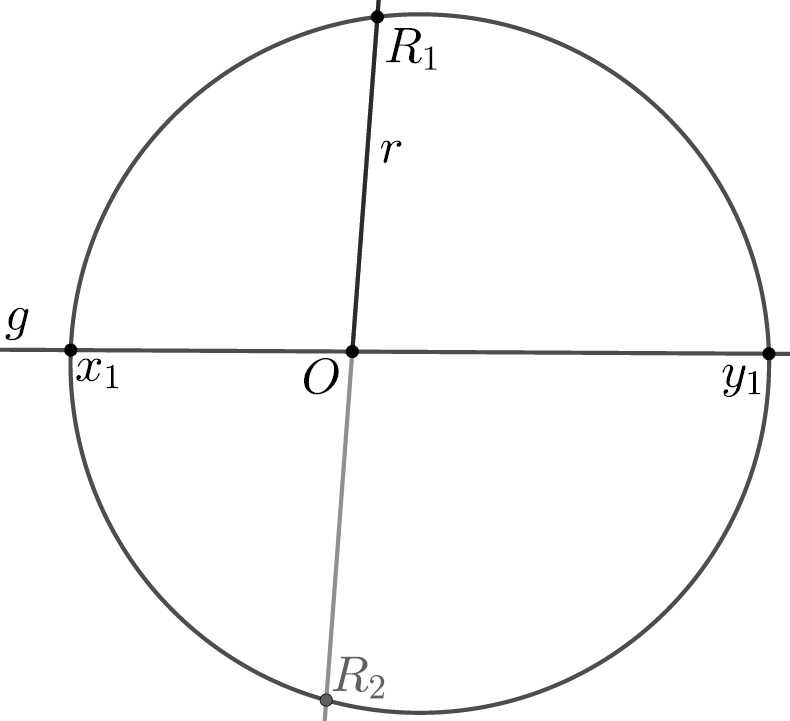}
\hspace{.03\textwidth}
\includegraphics[width=.30\textwidth]{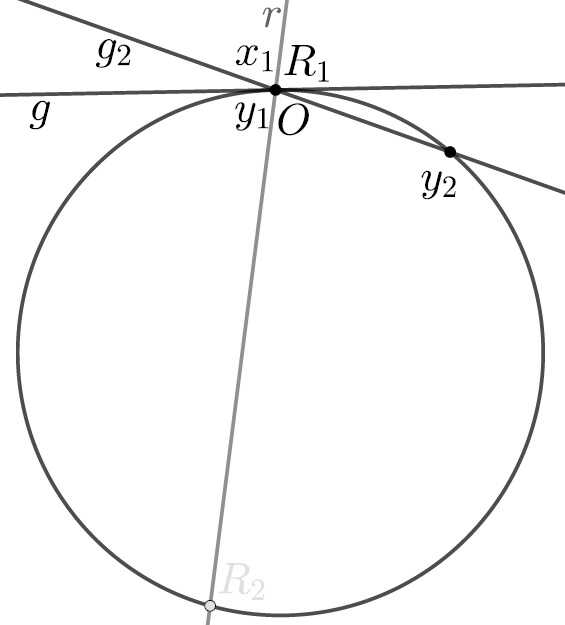}
\caption{Construction of a pair $x_1$, $y_1$ of an elliptic and parabolic involution on a line. (a) Einstein's second sketch on AEA~62-785r. Reproduced with permission from the Albert Einstein Archives. \copyright The Hebrew University of Jerusalem, Israel. Digital image photographed by Ardon Bar Hama. (b) Elliptic involution with no real invariant points; (c) Parabolic involution with one invariant point.}
\label{fig:785_2}
\end{figure}

Einstein concluded that setting $x=y$ gives two roots that can either be real or imaginary.\footnote{He wrote: ``Für $x=y$ zwei Wurzeln. Können reel[l] oder imaginär sein''.} We will see in \cref{fig:785_1,fig:785_2} that Einstein investigated both cases by drawing sketches.

Underneath a horizontal line, Einstein wrote down
\begin{align}
\label{eq:62785inv_3}
a : b : c = 
\begin{vmatrix}
x_1 + y_1 & 1 \\
x_2 + y_2 & 1 
\end{vmatrix}
\hspace{.06cm} \mathpunct{:} \hspace{.06cm}
\begin{vmatrix}
1 & x_1  y_1 \\
1 & x_2 y_2 
\end{vmatrix}
\hspace{.06cm} \mathpunct{:} \hspace{.06cm}
\begin{vmatrix}
x_1 y_1 & x_1 + y_1\\
x_2 y_2 & x_2 + y_2
\end{vmatrix}
.
\end{align}
As we already saw, an involution is determined by any two pairs $x_1$, $y_1$ and $x_2$, $y_2$ \cite[45]{coxeter_1987}. It can easily be shown that in this case, the real numbers $a$, $b$, and $c$ can be written as shown in \cref{eq:62785inv_3} when reading the colons not as divisions or homogeneous coordinates but simply as a symbol for separation of corresponding terms \cite[74]{blaschke_1954}.

As in \cref{eq:62785inv_1_2}, Einstein now considered $y$ to be the corresponding point to the point at infinity ($x = \infty$) yielding 
\begin{align}
y = - \frac{b}{a} = - \frac{
\begin{vmatrix}
				1 & x_1 x_1 \\
				1 & x_2 y_2
			\end{vmatrix}
			}
			{
			\begin{vmatrix}
				x_1 + y_1 & 1 \\
				x_2 + y_2 & 1
			\end{vmatrix}			
			}
	= + \frac{x_2 y_2 - x_1 y_1}{x_2 + y_2 - x_1 + y_1}
.			
\end{align}
As before, setting $y=0$ directly yields the constant of the involution.

On the same manuscript page, Einstein drew three sketches while one of these sketches obviously was not finished and crossed out. The first sketch is shown in \cref{fig:785_1}(a).

Interpreting the line $r$ in \cref{fig:785_1}(b) as the radical axis of the two circles defined by their intersections $R_1$ and $R_2$, it is 
\begin{align}
\overline{Ox_1} \cdot \overline{Oy_1} = \overline{Ox_2} \cdot \overline{Oy_2}.
\end{align}
This implies that any circle through $R_1$ and $R_2$ determines a pair  $x_i$, $y_i$ of the involution on line $g$ with center $O$ by the intersections of this circle with $g$ corresponding to \cref{eq:62785inv_2}. The two gray circles in \cref{fig:785_1}(b) are the only circles that pass through $R_1$ and $R_2$ and only touch line $g$. These two circles determine the two invariant points $S_1$ and $S_2$ of the (hyperbolic) involution, which were marked by Einstein. By \cref{eq:62785inv_2}, the distances between the invariant points and the center, respectively, are equal. Einstein indicated this by drawing a half-circle through the two invariant points with center $O$.

We summarize that Einstein constructed two pairs $x_1$, $y_1$ and $x_2$, $y_2$ of an involution with center $O$ as well as two real invariant points, clearly connected to the calculation above. As Einstein indicated in his calculation, the two invariant points could also be imaginary, leading to an elliptic involution. This case was considered by Einstein in \cref{fig:785_2}(a), titled ``In the imaginary case''\footnote{``Im imaginären Falle''.} by Einstein himself.

When moving up line $g$ in \cref{fig:785_1}(b) such that it comes to lie between points $R_1$ and $R_2$, we arrive at the situation in \cref{fig:785_2}(a), where Einstein did not draw the right circle anymore.\footnote{The transitions are visualized in the video sequence ``62785r.mp4''.} Clearly, no circle through $R_1$ and $R_2$ exists that only touches line $g$, since all circles passing through $R_1$ and $R_2$ need to cross the line $g$ twice. Einstein did not draw point $R_2$. However, by his notation $x_1$, $O$, and $y_1$ it becomes clear that the pair $x_1$, $y_1$ is separated by the center $O$ corresponding to an involution where line $g$ passes between $R_1$ and $R_2$.\footnote{In analogy to \cref{fig:785_1}(a) and the calculation, we argue that $x_1$, $y_1$ indeed is a pair of the involution and $O$ is the center.} In this case, the involution is elliptic, where no real invariant points exist. As we learn from Grossmann's lecture notes on Projective Geometry held by Fiedler and attended by Einstein, the difference between hyperbolic and elliptic involutions was taught by Fiedler as we find it on this manuscript page \cite[71]{grossmann_1897}: 
\begin{quote}
{\small [...] but those can be either real, then they are double points of the considered hyperbolic involution, or they are imaginary, such that the involution is elliptic.\footnote{``[...] diese können aber entw. reell sein, dann sind es die Doppelpunkte der betr. hyperbolischen Involution, od. sie sind imaginär, so ist jene Involution elliptisch.'' In this passage, the term \textit{Doppelpunkt} was used as it might be the case in the Scratch Notebook, see \cref{foot:doppelpunkt}.}
}
\end{quote} 
Corresponding sketches can also be found in the lecture notes of Grossmann's own lecture \cite[pp. 17.2 and 18.1]{grossmann_1907}.\footnote{See also \cref{foot:grossmann_1,foot:grossmann_2}.}

As we see in \cref{fig:785_2}(a), Einstein drew more lines as accounted for by our interpretation. Some of these lines are shown in \cref{fig:785_2}(c). Here, the center of the involution falls together with the point $R_1$. We get to this situation when moving up line $g$ in \cref{fig:785_2}(b). Thus, each circle that passes through $R_1$ and $R_2$ passes through the center of the involution, which is why every point on the line $g$ is a corresponding point to $O$. In particular, the circle drawn in \cref{fig:785_2}(c) determines the only invariant point of this so called parabolic involution, such that $x_1$ and $x_2$ fall together.

Instead of drawing another circle through $R_1$ and $R_2$ in order to determine a second pair of the involution, we can also alter line $g$. See for example line $g_2$ that passes through the center $O$, too. Point $y_2$ now corresponds to $O$ as well as to any other point on the line.\footnote{Einstein did not draw point $R_2$. This might be the reason why he considered another line $g_2$ instead of drawing a second circle in order to investigate parabolic involutions. We note that the involution on $g$ is not the same involution as on $g_2$.} It is likely that Einstein first wanted to draw the case of the parabolic involution in a third sketch, but then added it to the second sketch after crossing out the third sketch.\footnote{Another interpretation of this sketch is given in \cref{sec:alternatives}, where the point at infinity is an invariant point.}

We summarize that Einstein investigated hyperbolic, elliptic, and parabolic involutions on lines both algebraically and geometrically. The same involutions on conics instead of lines were investigated on manuscript page AEA~62-787r as we saw in \cref{subsec:62787r}.

\section{Further Sketches on AEA~124-446 and AEA~62-789}
\label{sec:further}

\begin{figure}[htbp]
\centering
\includegraphics[width=.30\textwidth]{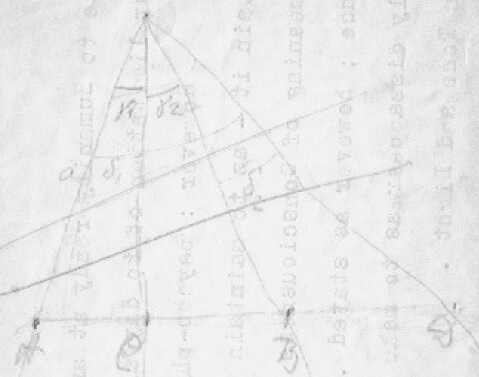}
\hspace{.03\textwidth}
\includegraphics[width=.30\textwidth]{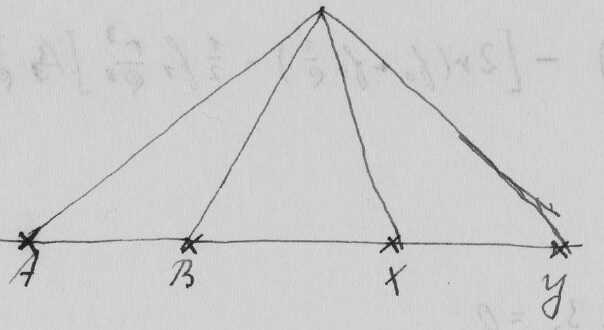}
\hspace{.03\textwidth}
\includegraphics[width=.30\textwidth]{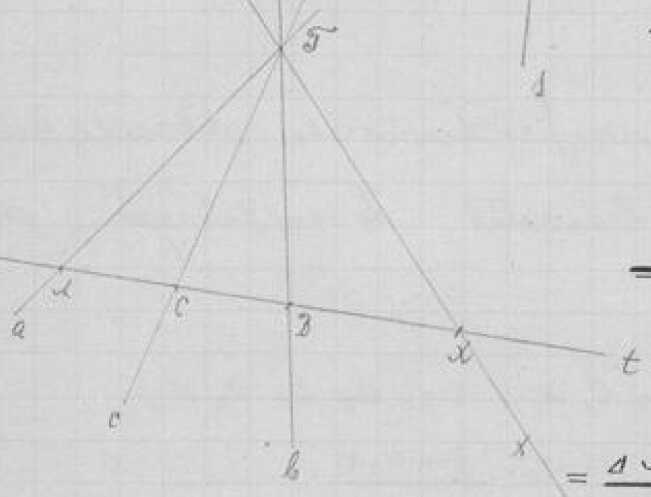}
\caption{Similar sketches on projective geometry on manuscript pages AEA~124-446 and AEA~62-789r as well as in Grossmann's lecture notes. (a) Einstein's first sketch on the manuscript page AEA~124-446; (b) Einstein's third sketch on manuscript page AEA~62-789; (c) Sketch from Grossmann's lectures notes on ``Darstellende Geometrie I'', original located at ETH. a) and b) are reproduced with permission from the Albert Einstein Archives. \copyright The Hebrew University of Jerusalem, Israel. Digital image photographed by Ardon Bar Hama.}\label{fig:124446_1}
\end{figure}

The first sketch on manuscript page AEA~124-446 has similarities to the third sketch on manuscript page AEA~62-789. Both sketches are shown in \cref{fig:124446_1}. As it was the case in \cref{subsec:general}, these are very generic sketches that appear in different contexts as perspectivities, cross ratios \cite[106]{enriques_1915}, involutions \cite[62]{coxeter_1987}, imaginary elements \cite[139]{dowling_1917}, or self-conjugate points \cite[61]{coxeter_1987}.

On AEA~124-446, Einstein wrote down
\begin{align}
\label{eq:124446_1}
\frac{AC}{BC} : \frac{AD}{BD} = \frac{BC}{AC} : \frac{BD}{AD}
\end{align}
and concluded
\begin{align}
\frac{AC^2}{BC^2} : \frac{AD^2}{BD^2} = 1.
\end{align}
We also find the expression
\begin{align}
\frac{\sin{\gamma_1}}{\sin{\gamma_2}} : \frac{\sin{\delta_1}}{\sin{\delta_2}},
\end{align}
which is equal to the expressions in \cref{eq:124446_1}. Einstein canceled certain quantities as well as added some squares in retrospect. The relations above can easily be proven as in \cite[105-110]{enriques_1915} in the context of cross ratios. By Grossmann's lecture notes, we know that Einstein already learned these relations in the winter semester 1896/97 when attending Fiedler's lecture ``Darstellende Geometrie I'', see \cite[38]{grossmann_1896} and \cite[363]{cpae1}. The corresponding sketch is shown in \cref{fig:124446_1}(c). The sketch on AEA~62-789 in \cref{fig:124446_1}(b) is very similar to these sketches, however, we do not find any related calculation on this manuscript page.

We also found a sketch on the reverse side of the letter AEA~6-250. It is remarkable that on this page, an equation is written down that is equivalent to an equation on AEA~62-789r which contains sketches on projective geometry as well, see \cref{subsec:general}.

\section{Alternative Interpretation}
\label{sec:alternatives}

We already saw that the center of an involution has a special meaning as it determines pairs of the involution. By our interpretation, Einstein did not draw the center when considering Pascal's theorem and its derivatives (except for \cref{fig:62787r_4}, where it falls together with a vertex). We can entertain different interpretations such that the center of the involution can be identified explicitly. In that case, however, certain lines drawn by Einstein become meaningless.

Einstein's sketches in figures \ref{fig:notebook_5}(b) and \ref{fig:62787r_3}(b), for instance, could also be interpreted as an involution that interchanges points $A$, $C$ and $D$, $F$, namely $ACD \barwedge CAF$.\footnote{In the sections above, we considered the involution that interchanges points $A$, $D$ and $C$, $F$.} In this case, the point $S_1$ becomes the center of the involution since the secants $FD$ and $AC$ passing through $S_1$ determine two pairs of the involution. As $S_1$ is an interior point, the involution is an elliptic one. In this case, Einstein did not draw the Pascal line\footnote{The axis of the involution then is the Pascal line of the hexagon $AADCCF$ and, thus, the line passing through the intersections of $AD$ with $CF$, $AF$ with $DC$, and the tangents in $A$ and $C$.} and, furthermore, the intersection $S_3$ of the tangents becomes meaningless.

Similarly, we could consider the involution $ADC \barwedge CFA$ instead of $ADC \barwedge DAF$ in \cref{fig:62787r_2}(b). In this case, the intersection $S_1$ is the center of the involution. This point might have been drawn by Einstein in the section where the paper had been torn off. However, the construction via the line passing through $S_1$ and $S_2$ would have been unusual, since he did not draw the line $EF$. The axis of the involution then crosses the conic twice, which is why the involution is hyperbolic. As before, in this case, the intersection $S_3$ of the tangents in $A$ and $D$ would be meaningless.

We can also interpret Einstein's sketch in \cref{fig:62787r_4} such that the center of the involution can be identified explicitly. Instead of looking at the parabolic involution $ACC \barwedge CAF$, we could consider the involution $FAD \barwedge DAF$ that interchanges the points $F$ and $D$ and let the point $A$ invariant. The center of the involution $P$ then is the intersection of the tangent in $A$ with the secant $DF$, which leads us to a hyperbolic involution. This interpretation, however, does not give a meaning to the line $PS_2$, either. Moreover, the intersection $S_3$ of the tangents in $A$ and $C$ would again be meaningless.

\subsection{Manuscript Page AEA~62-785r}

\begin{figure}[htbp]
\centering
\includegraphics[width=.45\textwidth]{files/62785_2.jpg}
\hspace{.03\textwidth}
\includegraphics[width=.45\textwidth]{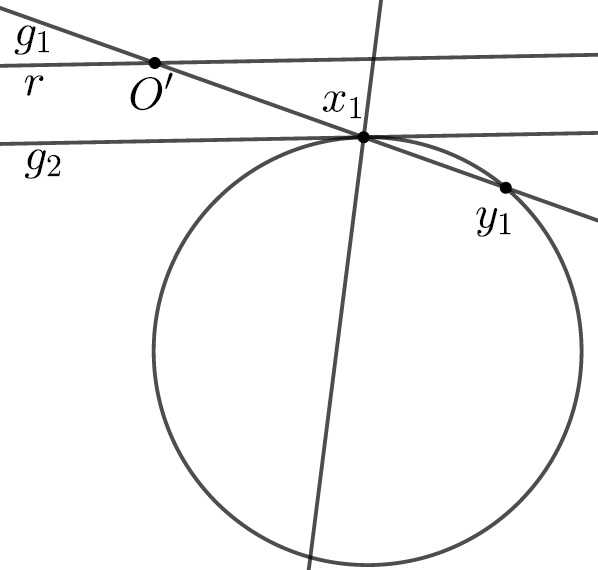}
\caption{Alternative Interpretation for Einstein's second sketch on AEA~62-785r. (a) Einstein's second sketch on AEA~62-785r. Reproduced with permission from the Albert Einstein Archives. \copyright The Hebrew University of Jerusalem, Israel. Digital image photographed by Ardon Bar Hama. (b) Alternative Interpretation where the point at infinity is an invariant point.}
\label{fig:785_3}
\end{figure}

In \cref{subsec:62785}, we argued that Einstein considered hyperbolic, elliptic and parabolic involutions on lines that correspond to both the calculations on the same page and considerations on the manuscript page AEA~62-787r. However, for the sketch in \cref{fig:785_2}(a) (see also \cref{fig:785_3}(a)) another plausible interpretation is possible.

We argued that Einstein implicitly assumed the point $R_2$ being on the circle such that the radical axis passes through the given circle. Then, we argued that the lines in \cref{fig:785_2}(c) show the parabolic involution. Instead of this, we can argue also as follows: A radical axis can also be defined when the two circles do not meet each other. Assuming that the second circle is above the circle drawn in \cref{fig:785_3}(a), the top horizontal line could be the radical axis (see the line $r$ in \cref{fig:785_3}(b)). Let $O'$ be the center of the involution on $g_1$, $x_1$ and $y_1$ is one corresponding pair. By rotating $g_1$ to $g_2$ around $x_1$, the point $O'$ goes to infinity. As $O'$ is the corresponding point to the point at infinity, it becomes an invariant point. In this case, $x_1$ becomes the second invariant point on the line $g_2$. This interpretation, however, does not fit to Einstein's calculation where he set the center of the involution as origin (real point). However, it would provide an interpretation for the third horizontal line in \cref{fig:785_3}(a).

\section{A Conjecture on the Purpose of Einstein's Ideas about Projective Involutions}
\label{sec:conjecture}

Given any one of the sketches individually, it might be well possible that Einstein would simply have been doodling. But the network of relations between the sketches and calculations, both in the Scratch Notebook and in the Princeton manuscripts, and the relations between the notebook and the manuscripts pages, and the fact that all of these sketches and calculations could be connected to the notion of an involution strongly suggest that Einstein was doing the constructions and calculations with a certain purpose in mind. But what could that purpose have been?

It might be tempting to associate any explicit considerations of projective geometry on Einstein's part with projective relativity of some form or other. Perhaps Einstein may have had in mind to interpret four-dimensional spacetime as some projected version of five-dimensional spacetime. Such association would be all the more justified by the fact that the relevant manuscript pages 62-785, 62-787, 62-789 all can be linked unambiguously to the Einstein-Bergmann correspondence around their five-dimensional generalization of Kaluza-Klein theory. However, we have not been able to establish any convincing link along these lines. 

Instead, we would like to express a different conjecture about the heuristics of Einstein's projective geometry sketches and calculations, as we have reconstructed them. We find equations that we can link to the correspondence between Einstein and Bergmann from summer 1938 on the three manuscript pages AEA~62-785r, 62-787r, and 62-789r. As it was stated in the follow-up of Einstein and Bergmann's publication, Einstein tried to find particle-like solutions within the framework of his five-dimensional approach during that time \citep{einstein_bargmann_bergmann_1941}.\footnote{See also \cite{dongen_2010} and \citet{SauerSchuetz2020Manuscript}. For further information about Einstein's five-dimensional approach, see \cite{dongen_2002}.}
We conjecture that in order to find these particle-like solutions, Einstein and Bergmann assumed a special parametrization of the five-dimensional metric, expressing periodicity with respect to the fifth coordinate, stationarity with respect to the (time-like) fourth coordinate and spatial (three-dimensional) spherical symmetry. Crucially, the spherical symmetry resulted in a dependence of the coefficients of the metric in terms of a radial coordinate $r$. Interpretation in terms of physical particles required regularity at the origin $r=0$ and a certain fall-off at (spatial) infinity $r\rightarrow\infty$. In order to satisfy the requirements, they considered power series expansions of the various coefficient functions. In this context, Einstein and Bergmann explicitly considered the mathematical problem to be that of an ``expansion around the infinite point'' (Einstein to Bergmann, Friday, i.e., probably 15 July 1938, AEA~6-242 and Bergmann to Einstein, 16 July 1938, AEA~6-264).\footnote{Bergmann dated his letter AEA~6-264 to 16 July, while he himself indicated at the end of the letter that he only sent it on 17 July. Einstein's letter AEA~6-242 can be dated to 15 July by surrounding correspondence.}

We therefore conjecture that the context of Einstein's considerations of involution in projective geometry is motivated by the wish to gain a better understanding of properties of a power series expansion around infinity. Typically, an expansion around infinity is done by a transformation $z\rightarrow 1/z$ and expanding around $0$. This mapping is a special case of the general involution equation (\ref{eq:involution}) or (\ref{eq:62785inv_1}).
Mapping the coordinates in such a way that the infinite point is mapped to a finite coordinate and by an involution, i.e. by a map that can be iterated to be inverted again would allow to investigate properties of the series expansion around infinity. The purpose of the considerations on involutions therefore may have been to consider a more sophisticated version of an expansion around infinity or to gain a geometric understanding of such a mapping. This conjecture will have to be tested against a more detailed reconstruction of Einstein's attempts to construct particle solutions in the Einstein-Bergmann framework.

It is intriguing to note that the consideration of projective involutions in the Scratch Notebook appears sandwiched in calculations of the magnification factor of gravitational lenses where Einstein had also found a diverging expression on the page immediately preceding it. However, while the sketches and calculations on the Princeton manuscripts all indicate clearly Einstein's interest in the infinite point, the infinite point does not appear explicitly in constructions of the Scratch Notebook. 

\section{Concluding Remarks}
\label{sec:conclusion}

Our analysis shows, we believe, that Einstein maintained an active command of basic notions of projective geometry throughout his life even though this field of mathematics did not play a prominent role in his physical theories. It is remarkable that very similar constructions and calculations dealing with involutions in projective geometry can be found that were written down some quarter century apart. If our conjecture is right, the constructions reveal a creativity on Einstein's part in working out solutions to unified field theory that have never made its way to published work. It will be interesting to further reconstruct the ideas documented in his research notes and working sheets.

\section*{Acknowledgements}
We wish to thank Diana Buchwald and the Einstein Papers Project at Caltech for their interest in and support of this project. We also thank Jeremy Gray and Klaus Volkert for helpful comments on an earlier draft of this paper.
Preliminary versions of this paper were presented at the 30.~\emph{Novembertagung} 2019 in Strasbourg  and at the BSHM meeting at Queen's College, Oxford, in February 2020. We thank the participants for discussion and their interest in our work.
We also wish to thank the Albert Einstein Archives, The Hebrew University of Jerusalem, for permission to quote from and publish facsimiles of unpublished documents.

\bibliographystyle{apalike}      
\bibliography{projgeom}   

\end{document}